\documentclass[letterpaper]{article} 
\usepackage{aaai2026}  
\usepackage{times}  
\usepackage{helvet}  
\usepackage{courier}  
\usepackage[hyphens]{url}  
\usepackage{graphicx} 
\usepackage{natbib}  
\usepackage{caption} 
\usepackage{array} 
\usepackage{algorithm}
\usepackage{algorithmic}
\usepackage{float}      
\usepackage{subcaption} 
\usepackage{amsmath} 
\usepackage{booktabs} 
\usepackage[svgnames]{xcolor}
\usepackage{tikz}
\usepackage{placeins}

\usepackage{algorithm}
\usepackage{algorithmic}

\usepackage{xcolor}

\usepackage{newfloat}
\usepackage{listings}
\floatstyle{ruled}
\newfloat{listing}{tb}{lst}{}
\floatname{listing}{Listing}

\title{The Bureaucratization of the Internet: \\ Analyzing the Diffusion of Governance Regimes on Reddit (2011--2023)}

\author{
    Katherine Van Koevering\textsuperscript{\rm 1},
    Yuanhao Liu\textsuperscript{\rm 2},
    Jon Kleinberg\textsuperscript{\rm 3}
}
\affiliations{
    \textsuperscript{\rm 1}Johns Hopkins Data Science and AI Institute\\
    \textsuperscript{\rm 2}Department of Sociology, Johns Hopkins University\\
    \textsuperscript{\rm 3}Department of Computer Science, Cornell University\\
    kvankoe1@jhu.edu, yliu514@jhu.edu, kleinber@cs.cornell.edu
}

\begin{document}
\maketitle
\begin{abstract}
As online platforms scale, distributed communities must develop complex governance structures to maintain order. Reddit represents a unique experiment in this process: millions of subreddits manage their own digital commons, yet together they form an interconnected ecosystem of platform governance. Using a historical dataset of subreddit rules from 2011 to 2023, we apply computational methods to characterize that ecosystem and identify mechanisms driving change. We identify seven distinct governance regimes and document a platform-wide drift toward irreversible bureaucratization operating through two reinforcing dynamics: a ratchet effect among existing communities and a cohort shift in which new communities increasingly launch with pre-packaged regulatory frameworks. Using Dyadic Event History Analysis, we find that community scale is the dominant and universal predictor of rule adoption, independent of network exposure. Diffusion flows through two complementary channels: normative transmission via shared moderator networks and mimetic transmission among topically similar communities. However, mimetic copying is selective: it drives adoption of content rules addressing shared topical challenges while communities resist copying operational and behavioral rules. Additionally, communities with overlapping user bases differentiate rather than converge, consistent with ecological accounts of niche competition, and prestige-based contagion is firmly rejected. Together, these findings reveal decentralized platform governance as a stratified ecosystem in which lateral coordination, scale pressure, structural inertia, cohort effects, and episodic platform coercion jointly produce irreversible formalization with direct implications for platform management.
\end{abstract}

\section{Introduction}
Content regulation is the hidden infrastructure of the modern internet \cite{roberts2019behind}. While platforms like Facebook and X (Twitter) rely on centralized enforcement, Reddit represents a unique experiment in \textit{distributed governance}. Millions of distinct communities (subreddits) manage their own digital commons \cite{ostrom1990governing}, enforcing local norms that create a patchwork of ``micro-nations'' with varying degrees of freedom. Together, these communities form an ecosystem of governance connected by shared moderators, topical affinities, and the authority of a single platform sovereign. We approach this ecosystem through three guiding questions: (1) What governance regimes exist and how stable are they? (2) What mechanisms drive governance change? (3) What are the vectors of diffusion --- and do they resemble the prestige-based contagion processes documented in other institutional fields?

To identify common regimes, we analyze a historical dataset of subreddit rules dating back to 2010. Using Latent Class Analysis (LCA), we cluster rule patterns into seven regimes. The changing prevalence of these regimes reveals a platform-wide trajectory of irreversible bureaucratization. This drift operates at two levels simultaneously: existing communities are pulled toward heavier regulation through a ratchet-like dynamic in which governance expands easily but rarely contracts, while new communities increasingly launch with pre-packaged regulatory frameworks rather than building from scratch, reflecting the ambient institutional expectations of a maturing platform.

Rule stability itself is heterogeneous. Foundational behavioral norms --- rules against harassment and off-topic content --- are widely shared and rarely revised once codified. By contrast, meta-rules governing enforcement mechanics are the most volatile, with month-over-month edits outnumbering rule-instances themselves, suggesting that communities struggle less with \textit{what} to regulate than with \textit{how} to enforce. When major changes do occur, they are nearly evenly split between total ruleset rewrites and wholesale creation or deletion of rulesets. Three major environmental shocks each reshaped the ecosystem, but in distinct ways: coercive isomorphism does not simply amplify diffusion uniformly, but can suppress peer-to-peer rule innovation when the platform absorbs the regulatory burden directly \cite{caplan2018isomorphism}.

Finally, using Dyadic Event History Analysis, we isolate the vectors of change. Community scale is the dominant and universal predictor of rule adoption, operating independently of network exposure --- growing communities bureaucratize whether or not their peers are doing so. Diffusion itself flows through two complementary channels: \textit{normative transmission} via shared moderator networks, in which moderators act as institutional brokers porting governance templates across their portfolios, and \textit{mimetic transmission} among topically similar communities, which copy each other's foundational rulebooks under uncertainty. Critically, however, mimetic copying is not uniform --- communities resist copying operational and behavioral rules from similar peers. In practice, communities with overlapping user bases push toward governance complementarity rather than mimicry, consistent with ecological accounts of niche competition \cite{carroll1985concentration}. Prestige-based contagion --- the model in which elite or high-traffic subreddits function as platform-wide broadcasters --- is firmly rejected. In its place, volunteer moderators emerge as a \textit{professional class} whose lateral coordination quietly standardizes the internet's regulatory infrastructure.

\subsection{Related Work}
This work situates our contribution within a broader landscape of community-authored rule regimes, governance tooling, and policy evolution. Decentralized, volunteer-driven governance is common across the internet, enabling diverse strategies and enforcement but breaking down when users or governors behave badly. Tendencies toward governance formation from nothing and solidification of power appear repeatedly in the literature---reflecting our own findings.
\paragraph{Rules Online}
Reddit devolves governance to subreddit-level institutions where volunteer moderators author rules and configure sociotechnical tools. Studies of AutoModerator document how community rules are operationalized and motivate improved human–machine coordination \cite{jhaver2019human}; interview research links moderator roles, tasks, and rule development across platforms \cite{seering2019moderator}; and log analyses highlight contested enforcement and the limits of automation in ``gray area'' cases \cite{alipour2026gray}. Topic-specific analyses show rapid diffusion of new rule domains such as AI-generated content policies \cite{lloyd2025ai}. Causal evaluations of interventions on r/The\_Donald report mixed but significant effects on user behavior \cite{trujillo2022make}, and field experiments show light-touch norm reminders improve conduct \cite{matias2019preventing}.
Descriptive work characterizes subreddit rule regimes at scale and over time. \citet{fiesler2018reddit} conduct a large-scale content analysis of rule pages, proposing a taxonomy of content, behavior, and procedural rules and documenting heterogeneity in scope, tone, and specificity, including template reuse and enforcement linkages. Longitudinal studies extend this via archived snapshots: \citet{fang2023shaping} assemble three-year timelines for 967 communities, quantifying edit operations and revealing heterogeneous tempos, topical diffusion, and event-driven updates; \citet{reddy2023evolution} similarly document clarifications, expansions, and retirements across 467 communities over 1.5 years. We extend this literature by classifying seven governance regimes and modeling multi-year transitions among them.
Wikipedia's decentralized governance has been widely studied, with work documenting policy growth, bureaucratic tendencies \citep{butler2008don,reagle2010good}, formal dispute resolution \citep{forte2008scaling}, and oligarchic centralization \citep{shaw2014oligarchy}. Beyond Wikipedia, locally authored rule sets appear across peer production \citep{keegan2017evolution}, gaming communities, decentralized social networks, and participatory governance infrastructures, with variation in community values cautioning against one-size-fits-all approaches.
\paragraph{Rules Offline}
Long before platform moderation, social scientists examined how organizations generate, diffuse, and stabilize rules. Neo-institutional theory holds that organizations converge on similar formal structures through coercive, mimetic, and normative pressures \citep{dimaggio1983iron,meyer1977institutionalized}; \citet{tolbert1983institutional} show that late adoption of civil-service reform was driven by legitimacy rather than functional need. Most directly relevant, \citet{march2000dynamics} reconstruct decades of rule changes at Stanford University, finding evolution through competence accumulation, problem-driven learning, and external shocks---the same mechanisms we observe across subreddits. \citet{alvesson2019neo} caution that isomorphism mechanisms risk losing analytical bite without sharper specification, which our dyadic design addresses; \citet{lounsbury2021new} document competing institutional logics that organizations selectively combine, which may explain why subreddits adopt overlapping but distinct rule repertoires under shared platform pressure.
Organizational ecology offers a counterweight: \citet{hannan1977population,hannan1984structural} argue that accumulated routines and reliability pressures make rule structures resistant to change; \citet{carroll1985concentration} formalizes this in resource-partitioning theory, showing generalists capturing the mass-market center push specialists toward peripheral niches, producing persistent diversity \citep{carroll2000demography}; and \citet{haveman2022power} integrates both perspectives, showing how rules, routines, and culture jointly constitute organizational power---paralleling our hypotheses about scale, calcification, and administrative diffusion.
These traditions developed from studies of durable, legally constituted organizations \citep{haveman2022power}. Online communities differ sharply: founding costs are near-zero \citep{benkler2006wealth}, membership is voluntary \citep{hirschman1970exit,teblunthuis2022nocommunity}, governance is volunteer-driven \citep{matias2019civic,seering2019moderator}, and the regulatory environment is set by a single platform rather than a state \citep{caplan2018isomorphism,gillespie2018custodians}. Whether isomorphic and ecological dynamics extend to this regime is the empirical question we take up.

\section{Theoretical Framework and Hypotheses}

Existing literature on rule adoption bifurcates into two domains: organizational scholars study formal bureaucracies where decision-makers follow institutional logics \cite{omahony2007emergence}, while network scholars treat adoption as viral contagion driven by peer exposure \cite{centola2010spread}. Reddit governance fits neither. Subreddits are not formal firms, but neither are they simple aggregates of individuals. We conceptualize them as components of a meta-organization---a network of legally autonomous communities that lack employment contracts but function as a collective system \cite{gulati2012meta}, facing structural constraints of coordination costs, niche maintenance, and external regulation \cite{engert2025self}, while relying on influence and social norms rather than fiat. We propose four mechanisms driving bureaucratization at the organizational and network levels.

Adopting a restrictive governance regime is costly, requiring enforcement labor and political capital \cite{seering2020reconsidering}. Standard contagion accounts are therefore insufficient; we look instead to Institutional Theory, which distinguishes \textit{Mimetic Isomorphism} (copying successful peers to reduce uncertainty) from \textit{Normative Isomorphism} (adoption driven by professionalization) \cite{dimaggio1983iron}. We suggest shared moderators act as institutional brokers whose administrative networks standardize rules across their portfolios \cite{johnson2014emergence, chen2021decentralized}.

\begin{quote}
\textbf{H1 (Network):} Rule adoption is driven primarily by
administrative overlap (shared moderators), rather than peer prestige or user overlap (social contagion).
\end{quote}
As communities grow, informal sanctions fail to scale and human moderation becomes unsustainable, necessitating formal policies \cite{he2019platform}. Organizational scale thus acts as an independent driver of governance, requiring bureaucratization to manage coordination costs independent of network effects \cite{engert2025self}.

\begin{quote}
\textbf{H2 (Scale):} Community size is positively associated with the rate of rule adoption, independent of network effects.
\end{quote}
Organizational Ecology posits that aging organizations develop reliable routines that resist change---structural inertia \cite{hannan1984structural}. While early-stage communities adapt rapidly \cite{faraj2011knowledge},
mature communities prioritize stability over innovation \cite{ransbotham2011membership}. 

\begin{quote}
\textbf{H3 (Calcification):} Community age is negatively associated with the rate of rule adoption; older communities are less likely to adopt new governance regimes than younger ones.
\end{quote}
Subreddits operate under the shadow of hierarchy of the platform owner. Just as industries self-regulate to forestall government intervention \cite{lenox2006role}, subreddits must adapt to platform-wide mandates. Major interventions---such as the 2020 deplatforming---function as exogenous shocks triggering coercive isomorphism, interrupting organic diffusion and forcing synchronization of governance to avoid sanctions \cite{gillespie2020expanding}.

\begin{quote}
\textbf{H4 (Shock):} Exogenous platform interventions significantly increase the rate of rule adoption across all communities, overriding internal diffusion mechanisms and driving homogenization.
\end{quote}

\section{Data}

\subsection{Data Collection and Cleaning}

\begin{figure}[H]
    \centering
    \includegraphics[width=\linewidth]{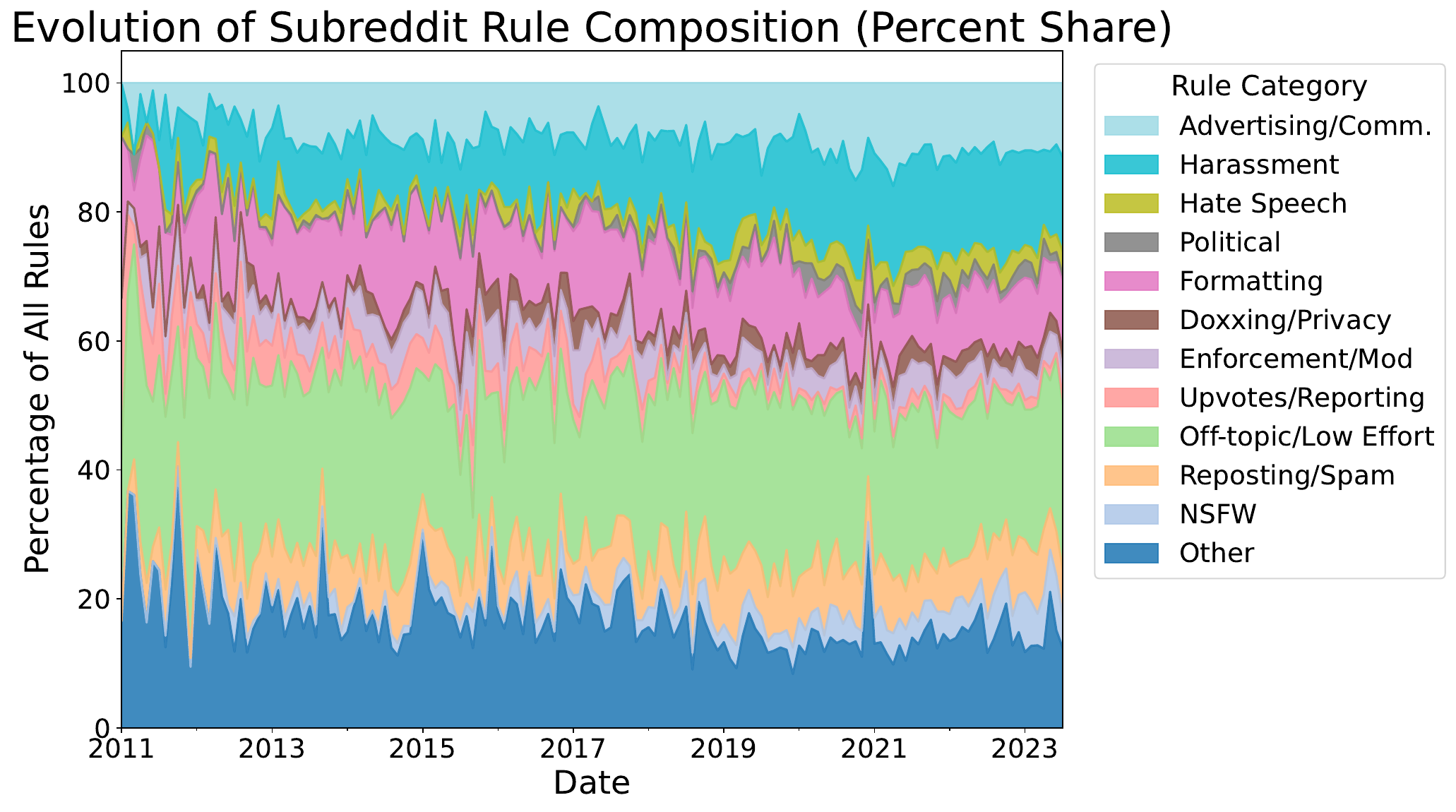}
    \caption{Evolution of Subreddit Rule Composition (Percent Share).}
    \label{fig:rule_evolution}
\end{figure}

We sourced subreddit rule pages from the Internet Archive's Wayback Machine,\footnote{\url{https://web.archive.org/}} collecting up to weekly snapshots of subreddit front pages from 2011 onward. For each snapshot, we extracted text from both the original moderator-editable sidebar and the standardized rules panel introduced by Reddit in later years (see Fig \ref{fig:rule_evolution}). Early rule text is noisy; we remove non-rule fragments (subreddit names, bare links, non-natural text) using custom regexes, and split remaining text into individual rules on newlines, numbers, or periods. Our final corpus comprises \textbf{10{,}183 rule-page snapshots from 2{,}281 subreddits}, spanning \textbf{January 2011 through July 2023}; the 1{,}985 subreddits with activity data for their archived months form the regime panel used in the diffusion models. The corpus contains \textbf{34{,}342 individual rule strings} (\textbf{24{,}932 unique} after exact-match deduplication). We also collected moderator records from subreddit about pages, yielding \textbf{138{,}310 moderator-snapshot records} covering \textbf{3{,}875 unique moderators}. Subreddit descriptions are drawn from a single recent Wayback snapshot per community. User overlap is measured using the Pushshift dataset \cite{baumgartner2020pushshift}, sampling active users one day per month per subreddit.

We construct a 12-category rule taxonomy building on prior work \cite{reddy2023evolution, fiesler2018reddit}. Starting from a combined prior taxonomy, we iteratively classified random subsets of rules, merging categories that shared more than 25\% of their rules until stability was achieved. We verified this against k-means clustering on embeddings of 25{,}000 rules ($k=10$), finding close correspondence with two exceptions---harassment/hate speech and advertising/doxxing--- which are empirically proximate but theoretically distinct and are retained as separate categories. Full category definitions and representative examples are provided in Appendix~\ref{sec:rule_categories} \footnote{Raw data available for download: 10.5281/zenodo.22210964 }.

\subsection{Automated Labeling and Validation}

Given the longitudinal scale of the dataset, we rely on Large Language Models to automate classification. To select the optimal model, three independent coders annotated a random sample of rules (mean pairwise inter-rater Cohen's $\kappa = 0.654$; Fleiss' $\kappa = 0.653$; nominal Krippendorff's $\alpha = 0.654$), and we evaluated three models against the majority-vote consensus. Fleiss' $\kappa$ and nominal Krippendorff's $\alpha$ are appropriate multi-rater measures for nominal labels \citep{geijer2025rater}.

As shown in Table~\ref{tab:model_validation}, Claude Sonnet 4.5 achieved the strongest alignment with human judgment (accuracy 67.00\%, Cohen's $\kappa = 0.620$), comparable to the inter-rater reliability among human coders (mean pairwise Cohen's $\kappa =0.654$; Fleiss' $\kappa =0.653$; nominal Krippendorff's $\alpha =0.654$). We therefore use Claude Sonnet 4.5 to label the full longitudinal dataset.

Some rules centered on Community Safety (such as \textbf{Harassment}, \textbf{Hate Speech}, and \textbf{Doxxing}) govern interpersonal user behavior and enforce boundaries on hostility. Interestingly, these rules range from formal, strictly legalistic language (``Do not post users' personal information'') to colloquial community maxims (``Rule 1: don't be a dick''). Others (such as \textbf{Off-topic}, \textbf{Politics}, \textbf{Advertising}, and \textbf{Spam}) dictate the thematic scope of the subreddit. Finally, categories like \textbf{Formatting}, \textbf{Meta/Enforce}, and \textbf{Upvotes} reveal how moderators attempt to standardize user submissions to make the subreddit readable and manageable.  Table \ref{tab:rule_examples} outlines all 12 categories utilized in our classification, alongside representative examples of rule text sampled directly from the dataset.

\section{Governance Regimes}

To understand the \textit{structure} of governance, we treat rules not as isolated variables but as components of a ``Regime.'' We employed Latent Class Analysis (LCA) \cite{mccutcheon1987latent} to cluster subreddits based on their rule portfolios, which identifies seven common regimes that describe governance strategies: Minimalist, Basic Civility, Content Curation, Safety \& Non-Commercial, Broad Regulation, Strict Regulation, and Other. We find that Reddit overall is moving away from laissez-faire governance and towards more formal regulation.

\subsection{Model Selection and Results}
We tested models ranging from $K=2$ to $K=10$ latent classes. The Bayesian Information Criterion (BIC) indicates that a 7-class solution provides the optimal balance of model fit and parsimony.

To identify these regimes, we employ Latent Class Analysis. We prefer LCA over K-means because it assigns probabilistic class membership and models binary inputs directly, avoiding the limitations of Euclidean distance metrics on sparse rule data (see Appendix~\ref{sec:model_specs}).

\begin{figure}[H]
    \centering
    \includegraphics[width=\linewidth]{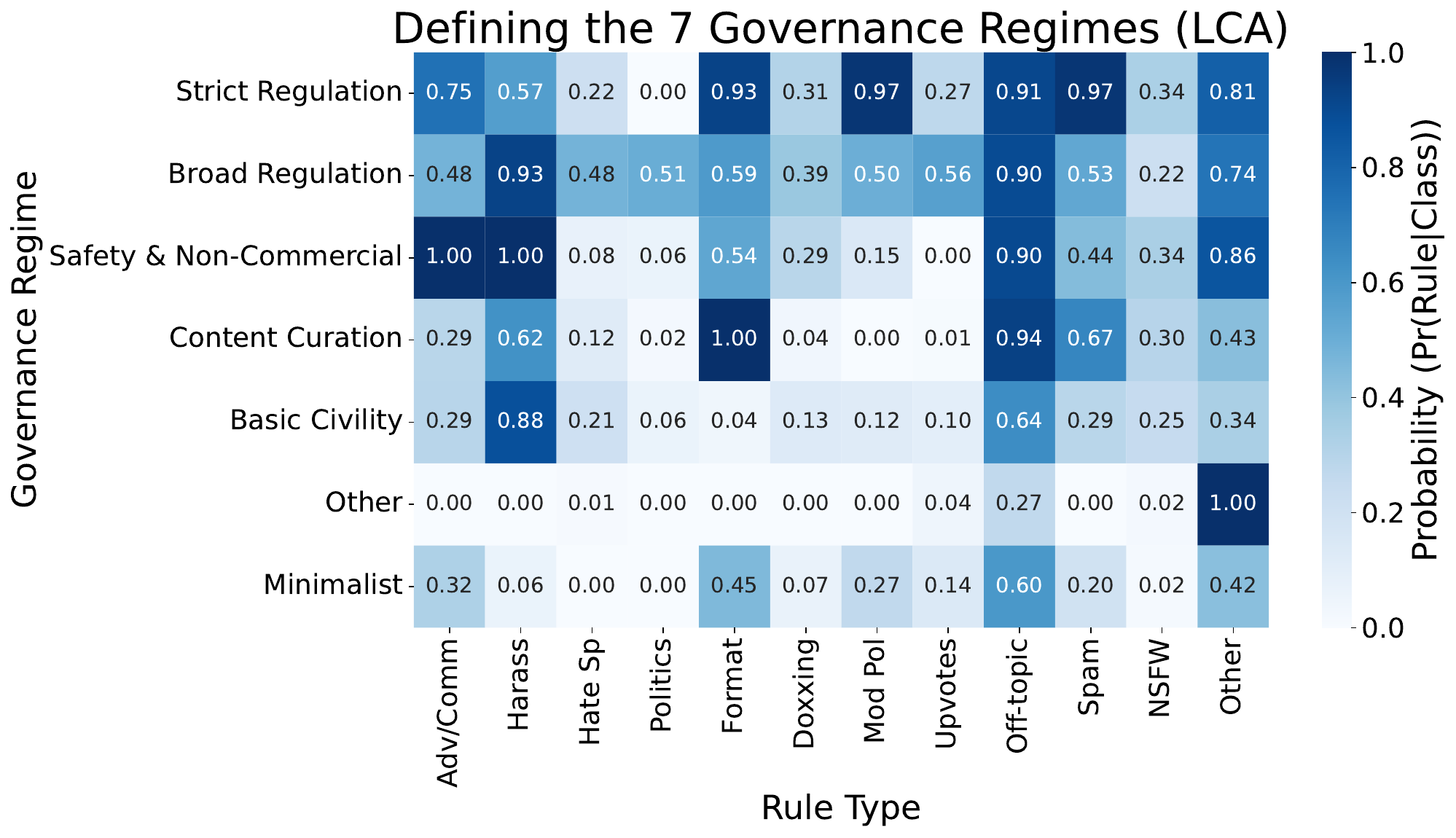}
    \caption{The Constitution of Seven Regimes (Conditional Probabilities).}
    \label{fig:regime_def}
\end{figure}

Based on the conditional probabilities of specific rules (Figure \ref{fig:regime_def}), we identified seven distinct regimes.
\begin{itemize}
    \item \textbf{Minimalists} have comparatively low probability of any rules. 
    \item \textbf{Content Curation} communities prioritize topic relevance and format ($Pr_{\text{Off-topic}}=0.94$, $Pr_{\text{Format}}=1.00$).
    \item  \textbf{Basic Civility} communities focus on tone ($Pr_{\text{Harassment}}=0.88$) with low regulation otherwise.
    \item \textbf{Safety \& Non-Commercial} communities add commercial regulation to civility concerns ($Pr_{\text{Adv/Comm}}=1.00$, $Pr_{\text{Harassment}}=1.00$, $Pr_{\text{Off-topic}}=0.90$).
    \item \textbf{Broad Regulation} communities exhibit comprehensive coverage, with the highest probabilities of Hate Speech ($Pr=0.48$), Politics ($Pr=0.51$), Upvotes ($Pr=0.56$), and Doxxing ($Pr=0.39$).
    \item \textbf{Strict Regulation} communities almost always regulate Format ($Pr=0.93$), Mod Pol ($Pr=0.97$), Off-topic ($Pr=0.91$), and Spam ($Pr=0.97$).
    \item \textbf{Other} communities always have the Other category ($Pr=1.00$) but rarely the clearly defined rule types, reflecting community-specific or humorous norms.
\end{itemize}

\begin{figure}[H]
    \centering
    \includegraphics[width=\linewidth]{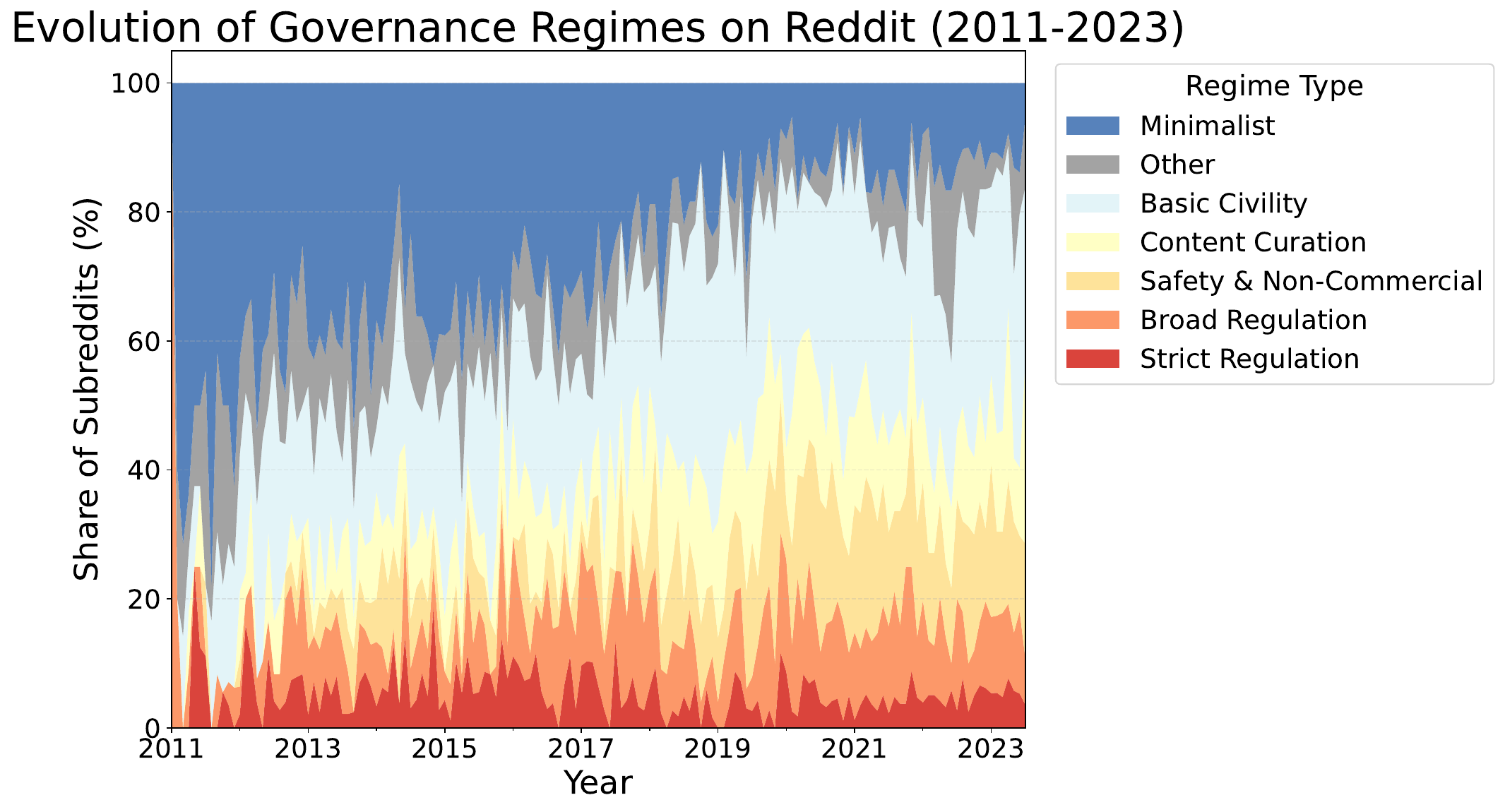}
    \caption{Regime Population Over Time.}
    \label{fig:regime_evo}
\end{figure}

To understand whether governance is a transient phase or a more permanent identity, we analyze the life histories of subreddits---the sequences of regimes each community occupies over time. We first convert each subreddit's regime sequence into a vector representation using TF-IDF, then cluster these vectors into six meta-trajectories. As shown in Figure~\ref{fig:path_dependence}, each trajectory cluster exhibits a clearly dominant regime, indicating substantial path dependence: subreddits tend to remain within their initial regime or drift toward a small number of related regimes rather than wandering freely across the regime space.


\subsection{Rule Stability and Change}

  We assess governance stability at two levels: the \emph{micro-level} volatility of rule categories, and the
  \emph{macro-level} shocks to entire community rulebooks.

  \subsubsection{Micro-Level: Volatility of Rule Categories}

  Let $c_{s,t,k}$ denote the number of rules in category $k$ for subreddit $s$ in month $t$. We define the volatility of
   category $k$ as
  $$V_k = \frac{\sum_{s,t} \lvert c_{s,t,k} - c_{s,t-1,k} \rvert}{\sum_{s,t} c_{s,t,k}},$$
 
The numerator counts every month-over-month change in category $k$, of which there are three types: creations ($0\rightarrow k$, when a subreddit first adds a rule of category $k$), deletions ($k \rightarrow 0$, when the last rule of category $k$ is removed), and quantity shifts within a persisting category (e.g., a subreddit moving from 2 to 3 Harassment rules). The denominator measures cumulative exposure: the total count of category-$k$ rule instances summed across every observed (subreddit, month) snapshot. A subreddit with three Off-topic rules observed in ten monthly snapshots thus contributes $30$ — not $10$ — to the exposure of Off-topic. Low $V_k$ marks ``sticky'' foundational rules that are written once and rarely revised; high $V_k$ marks ``slippery'' rules under constant revision.

  \begin{table}[ht]
  \centering
  \caption{Rule Volatility Ranking (Changes per Unit Exposure)}
  \label{tab:rule_volatility}
  \begin{tabular}{lrrr}
  \hline
  \textbf{Category} & \textbf{Changes} & \textbf{Exposure} & $V_k$ \\ \hline
  Enforcement/Meta        & 2{,}710 & 2{,}499  & 1.084 \\
  Other/Misc              & 8{,}420 & 8{,}618  & 0.977 \\
  Political Args          & 721     & 824      & 0.875 \\
  Hate Speech             & 1{,}237 & 1{,}508  & 0.820 \\
  Upvotes/Reporting       & 1{,}263 & 1{,}551  & 0.814 \\
  NSFW/Gore               & 1{,}575 & 2{,}006  & 0.785 \\
  Doxxing/Privacy         & 1{,}134 & 1{,}464  & 0.775 \\
  Formatting/Tags         & 4{,}596 & 6{,}175  & 0.744 \\
  Reposts/Spam            & 3{,}042 & 4{,}145  & 0.734 \\
  Advertising \& Promo    & 3{,}933 & 5{,}362  & 0.733 \\
  Harassment \& Civility  & 5{,}093 & 7{,}243  & 0.703 \\
  Off-topic/Low Effort    & 9{,}361 & 13{,}394 & 0.699 \\ \hline
  \end{tabular}
  \end{table}

  Table~\ref{tab:rule_volatility} reveals three tiers. \textbf{Foundational rules} ($V_k < 0.75$)—\textit{Off-topic/Low
  Effort} and \textit{Harassment \& Civility}—are the most prevalent yet the least volatile, suggesting basic behavioral
   etiquette is widely shared and rarely revised once codified. \textbf{Subjective rules} ($0.77 \leq V_k \leq
  0.88$)—\textit{Political Args}, \textit{Hate Speech}, \textit{NSFW/Gore}—churn more, as moderators iterate to patch  loopholes in contested gray areas. \textbf{Meta rules} ($V_k > 0.90$) are the most unstable: \textit{Enforcement/Meta} exceeds $1.0$, meaning month-over-month edits outnumber the rule-instances themselves. Communities, in other words,
  struggle less with \emph{what} to regulate than with the mechanics of \emph{how} to enforce.

  \subsubsection{Macro-Level: Anatomy of Governance Shocks}

  Beyond per-category churn, entire rulebooks occasionally undergo radical transformation. We compute the Jaccard
  distance between the multisets of rule categories in a subreddit's rulebook at $t$ and at its previous snapshot, and retain events with $\text{distance} \geq 0.5$, yielding
  $4{,}717$ shocks. Each is classified as:

  \begin{enumerate}
      \item \textbf{Creation} ($\emptyset \rightarrow$ rules): the first archived snapshot in which a subreddit's rulebook is observed.
      \item \textbf{Purge} (rules $\rightarrow \emptyset$): wholesale removal of the rulebook, typically signaling
  protest or moderator abandonment.
      \item \textbf{Replacement}: constitutional reform—the rulebook persists but its category composition changes by at
  least half (Jaccard distance $\geq 0.5$) between consecutive snapshots.
  \end{enumerate}


  In total, we find 2691 replacements, 1522 creations, and 504 purges. \textit{Replacement} dominates ($57\%$ of shocks), challenging an assumption of linear institutional inertia: established communities do not merely ossify, they periodically rewrite their constitutions. The prevalence of full-rulebook overhauls suggests that adaptation to platform mandates or demographic shifts often requires coordinated resets rather than incremental edits.

The appendix provides full detail on the temporal and distributional structure of macro-level governance shocks. We can see three platform events that we analyze later: the May 2018 redesign, which triggered a wave of rulebook formalizations; the January 6th, 2021, and a series of subreddit bans in 2020. Across all 2,691 Replacement events, the data support a ratchet interpretation, wherein more formal rule systems tend to persist and accumulate, whereas reversals are comparatively rare; the term is used by analogy to the retention of modifications in cumulative cultural change \citep{tennie2009ratchet}. 52\% of rewrites increase total rule count, the modal outcome is a net addition of one rule, and with the exception of a low-volume artifact in 2011 ($N=28$), 2018 is the only major year of net rule-shedding (mean $\Delta = -1.30$) - accounting for a significant portion of total rule-shedding and pulling down the average significantly. Direct moves between Broad and Strict Regulation are rare (30 escalations and 25 de-escalations), and in both directions the operational rules that define Strict Regulation shift sharply at the moment of reclassification (see Appendix~\ref{sec:shocks} for full results).

\subsection{The Pathways of Governance: Cohort Effects and Regime Transitions}

To understand how the governance ecosystem evolves, we must differentiate between two distinct phenomena: the shifting baseline of newly created communities (the cohort effect) and the evolutionary trajectories of existing communities (regime transitions).

\subsubsection{The Cohort Effect: The Shifting Baseline of New Communities}

Do communities always start as unregulated frontiers, or does the broader platform environment dictate their initial design? To test this, we isolated the first recorded governance regime for every subreddit in our dataset, grouping them by the year of their first archived rulebook.

\begin{figure}[tb]
    \centering
    \includegraphics[width=\linewidth]{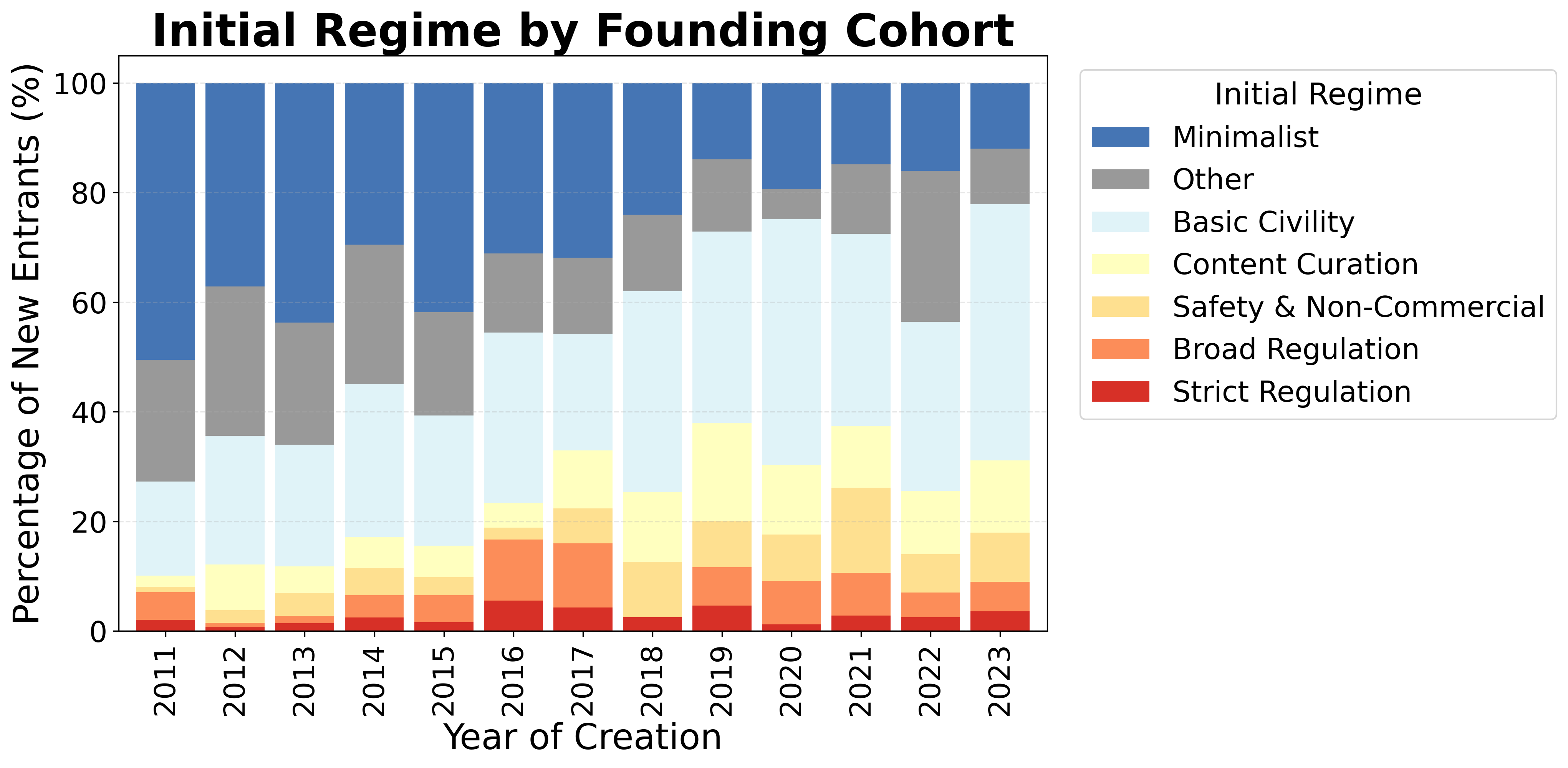}
    \caption{Cohort Effect: Initial Regime of New Subreddits by Year of First Archived Rulebook. This stacked bar chart illustrates the percentage of new entrants in each governance regime at their first archived snapshot.}
    \label{fig:cohort_effect}
\end{figure}

As illustrated in Figure \ref{fig:cohort_effect}, the ``default'' state of a new online community has fundamentally transformed over the past decade. In the early years of the platform (2011--2013), \textit{Minimalist} was the most common starting regime, accounting for 37--51\% of each cohort. The platform functioned largely as a laissez-faire frontier where governance was reactive. 

However, as the platform matured, we observe a stark decline in initial Minimalism. By 2016, and accelerating through 2020, new subreddits increasingly launched with pre-packaged regulatory frameworks, heavily favoring the \textit{Basic Civility} and \textit{Content Curation} regimes. This cohort effect suggests that new communities no longer start from scratch; the ambient expectations of the modern internet---now thick with institutional templates and platform-level mandates \citep{strang1993institutional, caplan2018isomorphism}---and the heightened coordination costs of immediate global visibility \citep{he2019platform} force administrators to deploy formal rulesets on day one, drawing on the rule repertoires of peer communities rather than improvising from zero \citep{kiene2025relational, frey2022governing}.

\subsubsection{Regime Transitions: The Gravity of Bureaucracy}

\begin{figure*}
\centering
\includegraphics[width=\linewidth]{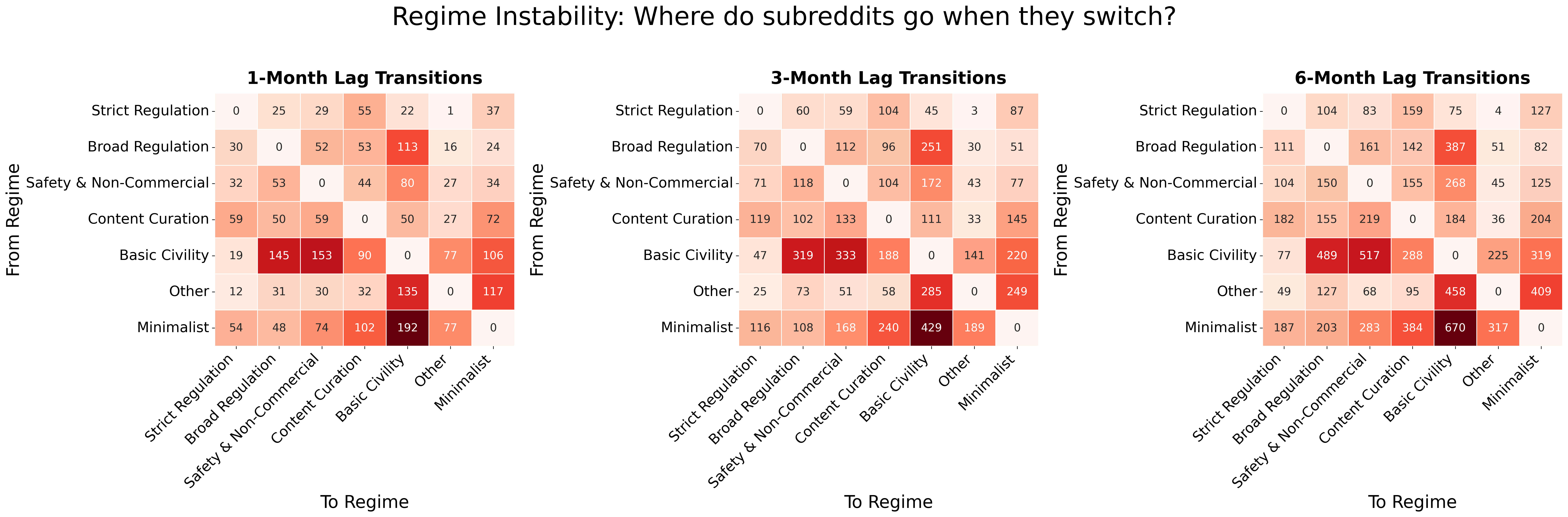}
\caption{Regime Instability and Transition Pathways. Heatmaps representing the volume of subreddit transitions between regimes across 1-month, 3-month, and 6-month lag windows. Darker shading indicates higher transition frequencies.}
\label{fig:transition_heatmaps}
\end{figure*}

While the cohort effect explains how communities are born, we must also examine where they go when their initial governance structures change. We measured regime transitions by tracking instances where a subreddit actively switched its governance class, analyzing the ``From-To'' pathways across 1-month, 3-month, and 6-month lag intervals.


The transition heatmaps (Figure~\ref{fig:transition_heatmaps}) reveal that governance evolution is rarely a random walk; rather, it exhibits distinct path dependence and a pronounced \emph{ratchet effect} --- the tendency for organizational structures, budgets, or regulations to expand easily in response to crises or growth, but to resist contraction once those immediate pressures subside. Several key patterns emerge from the transition matrices.

First, we observe a strong gravitational pull toward \textit{Basic Civility}, the most common destination overall. When communities abandon the \textit{Minimalist} regime, they flow most heavily into \textit{Basic Civility}.

Second, governance complexity moves upward more often than downward. Among one-month switches along the regime ladder, moves from lighter to heavier regulation outnumber the reverse by roughly three to two (1{,}160 vs.\ 796; e.g., 145 moves from \textit{Basic Civility} to \textit{Broad Regulation} versus 113 in the reverse direction). Downward moves are far from rare, but on balance communities that formalize their rules are more likely to add further structure than to repeal it.

Together, these analyses demonstrate that the bureaucratization of Reddit is driven by both external environmental pressures---forcing new communities to start with stricter rules in line with imprinting accounts of founding-period conditions \citep{stinchcombe1965social, marquis2013imprinting} and the platform-level coercive isomorphism that increasingly constrains new entrants \citep{caplan2018isomorphism}---and internal evolutionary gravity that pulls older communities toward increasingly complex regulatory regimes as rules accumulate through problem-driven learning and competence development \citep{march2000dynamics, frey2022governing, hwang2022rules}.

%

\section{Vectors of Change}

Having documented \emph{what} regimes exist and \emph{how often} they change, we now ask \emph{why} change occurs. We identify factors significantly correlated with changes in rules and regimes. In contrast to theories of viral social contagion, we find that diffusion flows primarily through professional administrative networks (shared moderators) and semantic homophily; community size acts as a primary catalyst for governance adoption, while older communities exhibit substantial structural inertia.

\subsection{Dyadic Time Event History Analysis}

To help determine \textit{why} these transitions occur, we constructed a dyadic dataset of all active subreddit pairs $(i, j)$ per month. We model the probability that target subreddit $j$ adopts a rule at time $t+1$, given that source subreddit $i$ already has it \cite{box1997dyadic}.

We utilize Dyadic Time Event History Analysis \cite{volden2006states, gilardi2008empirical}, shifting the unit of analysis to the directed pair $(i,j)$ at every time step $t$ to isolate specific transmission pathways and distinguish peer from elite influence (see Appendix~\ref{sec:model_specs}). Dyadic Time Event History Analysis treats each ordered source--target pair in each month as an observation, rather than treating a subreddit-month as the observation. It asks whether target $j$ adopts a rule after source $i$ already has it, allowing the model to distinguish relationships between communities from their individual characteristics.

We model the probability that Target subreddit $j$ adopts a rule at time $t$, \textit{conditional} on Source subreddit $i$ having already adopted it at time $t-1$. This approach offers two distinct methodological strengths:

\begin{enumerate}
    \item \textbf{Pathway Identification:} By interacting the attributes of the Source with the attributes of the pair (e.g., ``Do $i$ and $j$ share a moderator?''), we can statistically isolate specific vectors of transmission.
    \item \textbf{Relative Influence:} It allows us to test whether a subreddit is more sensitive to ``peers'' (similar size/age) or ``elites'' (larger/older communities), directly testing our hypotheses about mimetic isomorphism.
\end{enumerate}

We employ discrete-time logistic regression (full specification in Appendix~\ref{sec:model_specs}).

\subsection{Users, Moderators, and Topics}
We first compare three diffusion channels: social contagion (User Overlap), professional networks (Shared Moderators), and structural homophily (Topic Alignment). Table~\ref{tab:dyad_mechanisms} reports the corresponding log-odds coefficients.

These estimates speak directly to our Network Hypothesis (H1), which predicted that administrative overlap (shared moderators)---not viral contagion through shared users---would drive rule diffusion. The results support H1 in part and refine it in another.

The contagion premise is firmly rejected. \textbf{User Overlap} is negative across 14 of 19 outcomes and significantly so for seven of them (Broad Regulation, Basic Civility, Other-Regime, Spam, Other-Rule, Doxxing, and Harassment; Table~\ref{tab:dyad_mechanisms}); it is never significantly positive. Communities with overlapping user bases do not converge on identical governance; rather, they appear to actively differentiate, suggesting that overlapping populations create pressure for niche complementarity rather than mimicry---consistent with ecological accounts in which audience overlap pushes organizations toward niche specialization rather than convergence \citep{carroll1985concentration, hannan1977population}.

Consistent with the second component of H1, the \textbf{Professional Network (Shared Moderators)} channel emerges as a precise vector of administrative transmission. Coefficients are large and highly significant for the Off-topic ($\beta = 0.82$, $p<.001$) and Other-Rule ($\beta = 0.83$, $p<.001$) rules---exactly the categories where templated, low-controversy machinery (e.g., topic gating, generic catch-alls) is easiest to port across portfolios. At the regime level the channel is weaker: only adoption of the Minimalist regime is significantly associated with shared moderators ($\beta = 0.59$, $p<.05$). This pattern is consistent with normative isomorphism transmitted through professionalized brokers \citep{dimaggio1983iron, johnson2014emergence}. Where shared mods correlate negatively (Adv/Comm, Harassment), the underlying rules likely require subreddit-specific calibration that cannot be lifted wholesale.

However, H1 understated a third mechanism. \textbf{Topic Alignment} emerges as the \textit{strongest} positive vector overall, with the largest significant coefficients of any mechanism: Content Curation ($2.45$), Off-topic ($1.47$), Spam ($1.18$), Minimalist ($1.07$), and Basic Civility ($1.04$). Subreddits with semantically similar descriptions copy each other's foundational rulebooks. This is classical \textit{mimetic isomorphism} in the DiMaggio--Powell sense: communities facing similar functional challenges look to structural peers to resolve governance uncertainty. The revised picture is therefore that H1's rejection of viral contagion holds robustly, but rule diffusion runs through \textit{two} complementary channels rather than one---mimetic homophily among topical peers and normative transmission through shared administrative networks. This dual structure is consistent with the multiple-logics framing in recent neo-institutional theory \citep{lounsbury2021new}.

\subsection{Age and Size}
We next examine ecological characteristics—size and maturity—on the propensity to export rules (Source) versus to adopt them (Target). Table~\ref{tab:dyad_attributes} reports coefficients on the four attributes.

This dyadic specification offers a direct test of our Scale Hypothesis (H2) and Calcification Hypothesis (H3), while also bearing on the lateral-versus-top-down distinction embedded in H1.

\textbf{H2 is strongly and uniformly supported.} \textbf{Target Size} is the most universally positive predictor of rule adoption---significant for every outcome ($p<.001$ for all but Strict Regulation, where $p<.01$), with coefficients ranging from $0.20$ (Basic Civility) to $0.55$ (Formatting, Mod Pol). As communities grow, informal norms fail and they must import formal governance to maintain order, consistent with the functional-necessity logic developed by \citet{he2019platform}. Critically, this effect is independent of network exposure: target size remains the dominant predictor even when controlling for shared moderators, user overlap, and topic alignment. Scale, in other words, is not a proxy for network position---it is an autonomous driver of bureaucratization.

\textbf{Source Size} effects are small but mixed, and they do not reverse the lateral diffusion conclusion: Diffusion is lateral, not top-down. This finding sharpens H1's claim that prestige imitation is not the dominant mechanism, and resonates with classical resource-partitioning intuitions in which scale produces specialization rather than universal templates \citep{carroll1985concentration, carroll2000demography}.

\textbf{H3 receives partial, qualified support.} \textbf{Target Age} is significantly negative for the specialized and procedural rules (Mod Pol, Upvotes, Hate Speech, Spam, Other-Rule) and for two regimes (Broad Regulation and Safety \& Non-Commercial), consistent with the structural-inertia mechanism articulated by \citet{hannan1984structural}: older communities are slower to take on new procedural machinery. Yet Target Age is significantly \textit{positive} for the Off-topic rule and for three regimes (Strict Regulation, Content Curation, and Basic Civility), a pattern the simple calcification prediction did not anticipate. At the rule level, the evidence for calcification is therefore reasonably consistent; at the regime level it is mixed, and age does not uniformly slow movement into heavier regimes. All age coefficients are small (at most $0.013$ log-odds per month). \textbf{Source Age} is largely null, mirroring the Source Size pattern and reinforcing the lateral-diffusion conclusion.

\subsection{External Shocks}
Finally, we estimate the impact of platform-wide interventions on dyadic adoption rates. Table~\ref{tab:dyad_shocks} reports coefficients for time-period dummies corresponding to three major Reddit policy eras.

H4 predicted that exogenous platform interventions would raise rule adoption rates across communities, overriding internal diffusion mechanisms and driving homogenization. The results offer mixed but theoretically informative support, refining the simple amplification story into a more nuanced account of when shocks intensify versus displace peer diffusion.

The \textbf{Platform Reform (May 2018)} acted as the clearest catalyst for coercive isomorphism \citep{dimaggio1983iron, caplan2018isomorphism}: we observe significant spikes in Broad Regulation ($\beta = 1.69$, $p<.001$), Basic Civility ($\beta = 1.29$, $p<.001$), Politics ($\beta = 2.10$, $p<.001$), and Upvotes ($\beta = 2.32$, $p<.001$), alongside a sharp drop in adoption of the Minimalist regime ($\beta = -1.30$, $p<.001$). Communities scrambled to formalize rulebooks in response to sweeping platform changes, exactly the pattern H4 anticipates.

The \textbf{Deplatforming Event (June 2020)} produced a striking inversion that H4 in its simple form did not anticipate. While broad civility regimes diffused more widely (Basic Civility $+0.91$, Safety \& Non-Commercial $+1.29$, both $p<.001$), dyadic copying of the specific \textit{Hate Speech} ($-1.30$) and \textit{Harassment} ($-1.04$) rules turned sharply \textit{negative} (both $p<.001$). This is a substantive qualification: when the platform itself absorbs the regulatory burden through top-down enforcement, the local need for peer-to-peer copying of the specific mandated rule \textit{diminishes}. Coercive isomorphism does not always intensify diffusion---it can displace it. Communities reallocate rulemaking attention away from the mandated rule and toward the broader civility framing where they retain discretion.

The \textbf{2020 Election / Jan 6 (Jan 2021)} shock produced a broad contraction rather than hardening: no regime diffused significantly faster, adoption of Safety \& Non-Commercial ($-1.47$, $p<.001$) and Other-Regime ($-1.36$, $p<.05$) fell, and several niche rules contracted (NSFW $-1.06$, Adv/Comm $-0.36$, Other-Rule $-0.45$, all $p<.01$). Rather than homogenizing communities, this period slowed peer-to-peer adoption, a further qualification of the homogenization story implied by H4.

Taken together, the three shock periods support the core claim of H4---platform interventions reshape diffusion patterns---but reveal that the mechanism is not uniform amplification. Shocks redistribute diffusion across the rule space, sometimes synchronizing communities on broad regimes, sometimes slowing adoption overall, and suppressing peer-to-peer transmission of the specific rules the platform now enforces directly.

\subsection{Ecological Drivers of Rule Adoption: Shared Challenges and Niche Competition}

To test the ecological mechanisms driving governance diffusion, we deploy a two-part analytical strategy. First, we test the \textit{Shared Challenge Hypothesis} by comparing the main effect of topic similarity across distinct rule categories. Second, we test the \textit{Niche Competition Hypothesis} using a split-sample interaction model to determine if structural equivalence (i.e., occupying the exact same size niche) triggers differentiation inhibition.

To ensure stable estimation and allow for direct magnitude comparisons across our models, all continuous predictors are standardized (Z-scored). Furthermore, to isolate the true behavioral effects of similarity and overlap, all models control for structural confounds including source size, target size, community age, global rule prevalence, and time-period fixed effects.

\subsubsection{Test 1: The Shared Challenge Hypothesis}

\paragraph{Rule Categorization}
If communities adopt rules strategically to address specific environmental problems, the effect of topic similarity should vary strictly by the nature of the rule being adopted. We aggregate the twelve specific governance rules into three distinct theoretical categories based on the nature of the regulatory challenge they address: \textbf{Content Rules (Topic-Specific):} This category includes \textit{Hate Speech, Politics, Off-topic,} and \textit{NSFW}. These rules define the boundaries of acceptable discourse and filter the substantive material allowed within the community. \textbf{Technical Rules (Operational):} This category includes \textit{Formatting, Moderator Policy (Mod Pol), Upvotes,} and \textit{Spam}. These rules address the mechanical, structural, and administrative challenges of maintaining a functioning online community. They govern the "how" of platform interaction rather than the "what," making them relatively universal operational standards. \textbf{Social Rules (Behavioral Norms):} This category encompasses \textit{Harassment, Doxxing, Advertising/Commercial (Adv/Comm),} and \textit{Other}. These rules regulate interpersonal conduct and establish baseline behavioral norms between users. We theoretically distinguish these from Content rules: whereas Hate Speech rules regulate the ideological boundaries of acceptable \textit{ideas}, Harassment and Doxxing rules regulate abusive \textit{actions} directed at specific individuals—a universal social challenge that communities face regardless of their specific topic.

\paragraph{Analytical Approach}
We estimate separate dyadic logistic regression models for three rule categories (Technical, Content, and Social, with $N=4$ rules per category). The baseline model is specified as:

\begin{equation}
    \begin{split}
    \text{logit}(P(\text{Adopt}_{i,t+1})) = \beta_0 + \beta_1 \text{CosineSim} \\
    + \beta_2 \text{LogOverlap} + \mathbf{X}\gamma + \epsilon
    \end{split}
\end{equation}

Where $\mathbf{X}$ represents our vector of structural controls. The key parameter of interest is $\beta_1$.

\paragraph{Results and Interpretation (Panel A)}
The fully controlled subsample models reveal a stark divergence that aligns with the shared challenge prediction (see Figure \ref{fig:ecological_diffusion}, Panel A). 
\begin{itemize}
    \item \textbf{Content Rules:} Demonstrate a strong positive standardized effect ($\beta = 0.052$).
    \item \textbf{Technical Rules:} Demonstrate a negligible negative effect ($\beta = -0.002$).
    \item \textbf{Social Rules:} Demonstrate a moderate negative effect ($\beta = -0.030$).
\end{itemize}

Accounting for foundational structural differences reveals that similarity \textit{only} drives the diffusion of topic-specific rules. When communities share a topic, they actively adopt each other's Content rules to combat shared threats (e.g., overlapping user demographics bringing similar disruptive behaviors). However, communities exhibit strategic resistance toward universally applicable operational and behavioral rules, demonstrating that governance diffusion is a targeted ecological adaptation rather than blind mimicry.

\subsubsection{Test 2: The Niche Competition Hypothesis}

\paragraph{Analytical Approach}
We introduce an interaction term ($\text{LogOverlap} \times \text{CosineSim}$) and split our risk set into two structural groups based on the median size difference between the source and target communities: \textit{Peers} (same size niche) and \textit{Non-Peers} (different size niches). 

\paragraph{Results and Interpretation (Panel B)}
The interaction effects (Figure \ref{fig:ecological_diffusion}, Panel B) provide robust empirical support for the niche competition mechanism. 
\begin{itemize}
    \item \textbf{Non-Peers (Different Sizes):} Exhibit a strong positive interaction ($\beta = 0.044$). When communities occupy different ecological niches, high overlap and high similarity function synergistically to drive convergence. They are structurally complementary, allowing for the free flow of institutional practices.
    \item \textbf{Peers (Same Size):} The interaction effect is entirely neutralized and turns slightly negative ($\beta = -0.005$). 
\end{itemize}

This finding successfully isolates the differentiation inhibition effect. The pressure to coordinate with or copy successful peers is actively overridden by the strategic need to differentiate, but \textit{only} when communities are direct structural competitors fighting for the same user base.

\section{Conclusion}
This study offers a systematic account of how governance evolves across a large-scale decentralized platform. Analyzing subreddit rule histories from 2011 to 2023, we identify seven stable governance regimes and demonstrate that the platform's trajectory is one of irreversible bureaucratization---a drift from laissez-faire minimalism toward increasingly formalized, comprehensive rule systems. Critically, this drift is not accidental: it is structured by identifiable mechanisms operating at multiple levels simultaneously. Our dyadic event-history analysis allows us to evaluate each one against the four hypotheses we set out at the start, and we summarize those verdicts here.

\paragraph{H1 (Network) --- supported with refinement.} The data firmly reject the viral-contagion premise: communities with overlapping user bases do \textit{not} converge on similar governance, and elite or prestigious subreddits do not function as platform-wide broadcasters. In place of contagion, two complementary mechanisms drive diffusion. Shared moderators act as ``institutional brokers,'' porting templated, low-controversy machinery across their portfolios---normative transmission consistent with classical accounts of professionalized fields \citep{dimaggio1983iron}. Topical peers, identified through semantic embedding similarity, copy each other's foundational rulebooks---mimetic transmission consistent with structural-equivalence accounts of imitation under uncertainty. The original H1 framing---that administrative overlap is the singular dominant channel---is therefore qualified: mimetic homophily and normative transmission operate in tandem, aligning with the multiple-logics turn in recent neo-institutional theory \citep{lounsbury2021new}. The substantive contribution stands: volunteer moderators are not merely content enforcers but a \textit{professional class} whose lateral coordination quietly standardizes the internet's regulatory infrastructure.

\paragraph{H2 (Scale) --- strongly supported.} Target community size is the most universally positive predictor of rule adoption, significant for every outcome we examined and independent of network exposure. Growing communities face functional pressure to bureaucratize whether or not their peers are doing so, echoing the offline pattern in which scale and coordination cost drive the emergence of formal rule systems \citep{march2000dynamics, haveman2022power}. Source size, by contrast, is essentially null---reinforcing the lateral-diffusion conclusion of H1.

\paragraph{H3 (Calcification) --- partially supported.} Older communities are slower to adopt specialized and procedural rules (Mod Pol, Upvotes, Hate Speech, Spam), as the structural-inertia logic predicts \citep{hannan1984structural}. At the regime level, however, age effects are mixed: older communities are less likely to move into Broad Regulation and Safety \& Non-Commercial but more likely to move into Strict Regulation, Content Curation, and Basic Civility, so calcification does not uniformly block movement toward heavier regimes.

\paragraph{H4 (Shock) --- supported with mechanism refined.} Platform-level interventions reshape diffusion, but not by simple amplification. The 2018 redesign triggered broad coercive isomorphism across multiple regimes, matching H4 cleanly. The 2020 deplatforming, however, produced a paradoxical \textit{suppression} of peer-to-peer copying of the specific rules the platform now enforced directly---communities reallocated their rulemaking attention to civility framing where they retained discretion. The 2021 post-election period brought a general slowdown in peer-to-peer adoption, including of several niche rules and regimes. Coercive isomorphism, in short, is heterogeneous: top-down enforcement can either intensify or displace peer diffusion depending on whether the platform absorbs the regulatory burden directly \citep{caplan2018isomorphism}.

Together, these findings position decentralized platform governance as neither a flat diffusion field nor a top-down hierarchy, but a stratified ecosystem in which lateral mimicry among topical peers, normative coordination among moderator portfolios, scale-driven bureaucratization, age-graded inertia, and episodic platform coercion jointly produce the irreversible drift toward formalization that we document. The mechanisms identified offline by neo-institutional and ecological scholarship persist online, but the relative weight of each channel shifts: the absence of legal contracts and employment relations elevates lateral mimicry and reduces the influence of elite broadcasters, while the presence of a single platform sovereign concentrates coercive force in a way no offline analogue cleanly mirrors.

These findings carry implications for platform designers and policymakers. If governance diffuses through moderator networks and topical homophily rather than imitation of popular peers, then investing in moderator tooling, coordination infrastructure, and communities-of-practice across topical clusters is likely to have outsized effects on platform-wide governance quality. Conversely, top-down policy mandates risk homogenizing a diverse ecosystem in ways that eliminate the very variation that makes distributed governance valuable---and may, paradoxically, suppress the peer-to-peer rule innovation that helps communities address novel problems.

\subsection*{Limitations}

Several limitations bound our conclusions. First, our data derives from Wayback Machine snapshots, introducing uneven temporal coverage across subreddits; communities with fewer archived snapshots may have governance changes that are missed or imprecisely dated. Second, while we measure moderator overlap as a proxy for administrative network ties, we cannot directly observe \textit{how} moderators communicate or transfer practices --- the mechanism is inferred structurally rather than observed behaviorally. Third, our rule taxonomy, while grounded in prior work, reduces the rich heterogeneity of rule text to categorical indicators; two communities coded identically may enforce qualitatively different norms. Fourth, the seven-regime LCA solution depends on binary coding of rule presence; we adopt binary coding both for comparability with prior taxonomic work and because Bernoulli LCA is the standard fit for presence-only categories. Finally, our analysis is bounded to Reddit, a platform with a distinctive federated structure and an unusually active volunteer moderator class; generalization to platforms with different governance architectures --- such as Discord, Facebook, or Wikipedia --- should be made cautiously.

\FloatBarrier
\bibliography{references_ver4}

\raggedbottom 
\appendix

\section{Appendix}
\subsection{Evolution of Reddit Rules}
\begin{figure*}[t]
\centering
\includegraphics[width=1.0\linewidth]{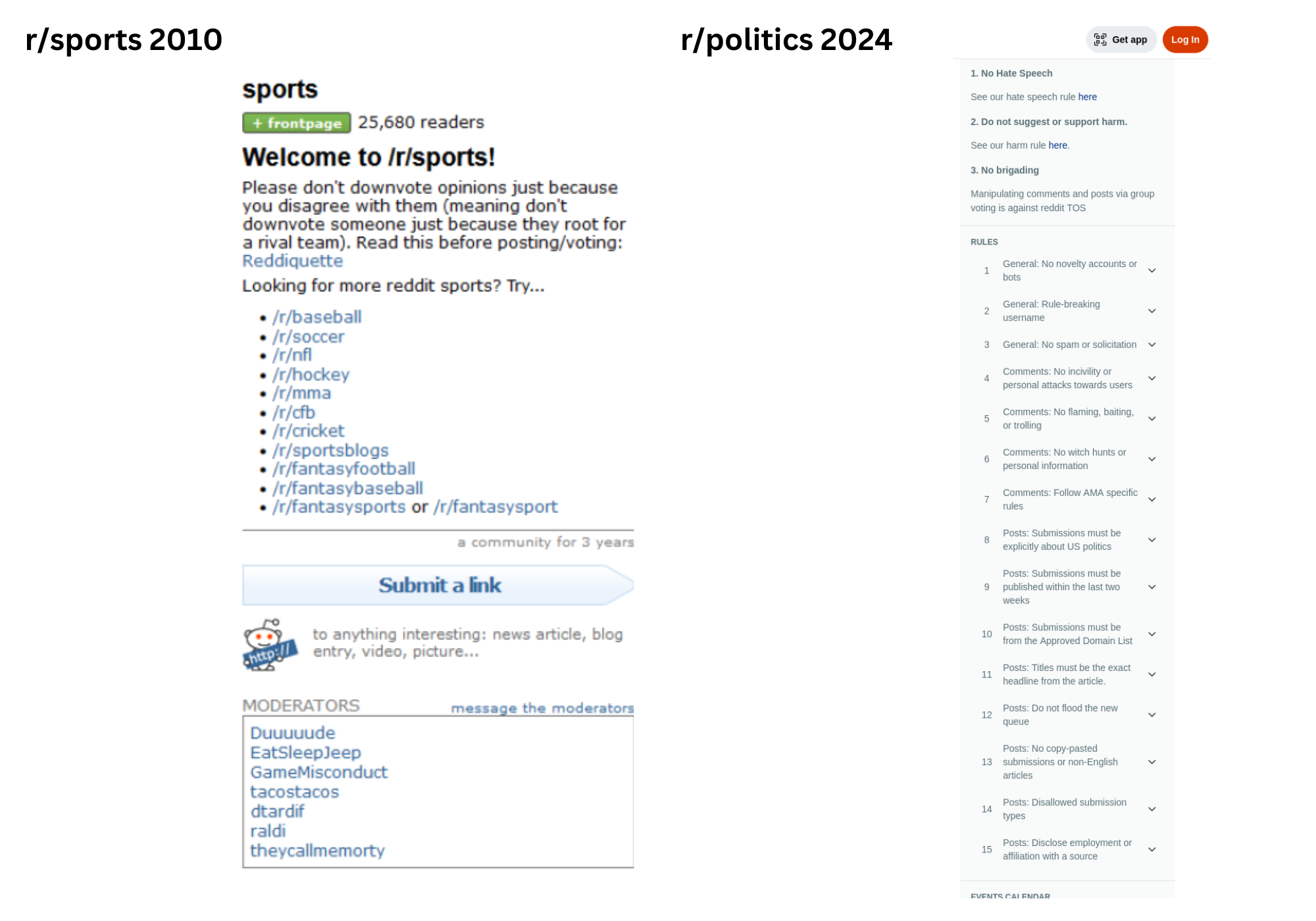}
\caption{\textbf{Evolution of subreddit rule structures over time.} The left panel shows r/sports in 2010, where the only visible rule is an informal guideline asking users not to downvote opinions they disagree with. This reflects an early stage of rule-making on Reddit, where norms were loosely defined and enforcement relied on community expectations rather than explicit moderation policies. In contrast, the right panel presents r/politics in 2024, where the rules are extensive, systematically categorized, and explicitly stated.}
\label{fig:politics_new}
\end{figure*}

Figure~\ref{fig:politics_new} grounds the quantitative analysis in a concrete qualitative contrast that motivates the entire paper. The transformation it illustrates is not cherry-picked: it reflects the aggregate trend documented in Figures~\ref{fig:rule_evolution} and~\ref{fig:regime_evo} in the main text. What the figure makes viscerally clear is that the signal we are measuring is visible in the rule page itself. The contrast between r/sports 2010 and r/politics 2024 is, in miniature, the bureaucratization thesis. It also demonstrates the difficulty of classifying early rules: rules were often part and parcel of a larger description, mission, or explanation of the subreddit. It took years to develop distinct rule-sets used in most prior work.

\subsection{Model Specifications \label{sec:model_specs}}

\subsubsection{Latent Class Analysis}

We employ Latent Class Analysis (LCA) to cluster subreddits based on
their rule portfolios. We prefer LCA over K-means for two reasons.
First, LCA assigns probabilistic class membership, handling the
ambiguity of governance where a subreddit may exhibit features of
multiple regimes. Second, standard Euclidean distance metrics fail on
sparse, high-dimensional binary rule data; LCA explicitly models the
conditional probabilities of binary inputs.

Formally, LCA assumes the observed association between binary rule
variables is explained by a categorical latent variable $C$ with $K$
classes. For a subreddit $i$ with a vector of observed rules
$\mathbf{y}_i = (y_{i1}, \ldots, y_{iJ})$, where $y_{ij} \in \{0,1\}$
indicates presence of rule $j \in \{1,\ldots,12\}$, the probability of
observing that vector is:

\begin{equation}
P(\mathbf{y}_i) = \sum_{k=1}^{K} P(C=k) \prod_{j=1}^{J} P(y_{ij} \mid C=k)
\end{equation}

where $P(C=k)$ is the prevalence of regime $k$, and $P(y_{ij} \mid
C=k)$ is the conditional probability of having rule $j$ given
membership in regime $k$. We tested models from $K=2$ to $K=10$; BIC
indicated $K=7$ as the optimal solution.

\subsubsection{Dyadic Event History Analysis}

To model the probability that target subreddit $j$ adopts a rule at
time $t$, conditional on source subreddit $i$ having already adopted it
at $t-1$, we employ discrete-time logistic regression. The log-odds of
adoption are modeled as:

\begin{equation}
\ln \left( \frac{p_{ijt}}{1 - p_{ijt}} \right) =
\alpha + \beta_1 \mathbf{N}_{ijt} + \beta_2 \mathbf{A}_{jt} +
\beta_3 \mathbf{S}_t + \epsilon
\end{equation}

\subsection{Categories \label{sec:rule_categories}}
Table~\ref{tab:rule_examples} presents the full 12-category rule taxonomy used throughout this analysis, with representative examples drawn directly from the dataset. Constructing this taxonomy required balancing two competing demands: categories must be distinct enough that the classifier can reliably separate them, but theoretically coherent enough to support the regime-level analysis. We validated this balance both through LLM classification agreement (Table~1 in the main text) and through k-means clustering on rule embeddings (k=12), finding that two pairs of categories --- harassment/hate speech and advertising/doxxing --- were empirically similar by embedding distance but theoretically distinct enough to retain as separate categories. The ``Other'' category warrants particular attention: it is a substantive regime marker, capturing community-specific content and humorous norms that resist classification into the eleven substantive categories, as well as text that is more descriptive and may not be a rule at all. As shown in Figure~\ref{fig:regime_def}, ``Other'' is the defining feature of the \textit{Other} regime ($Pr = 1.00$), indicating communities whose governance is so idiosyncratic that it cannot be assimilated into any of the six substantive regime types.

\begin{table*}
    \centering
    \renewcommand{\arraystretch}{1.3} 
    \begin{tabular}{>{\raggedright\bfseries}p{2.5cm} p{12cm}}
        \toprule
        Category & \textbf{Example Rule Texts} \\
        \midrule
        Advertising & 
        No self promotion or advertising services or asking for money for yourself or another.\\
        
        \midrule
        Harassment & 
        No brigading, antagonism, name calling or insults.
Be civil, be nice, reddiquette.\\
        
        \midrule
        Hate Speech & 
        No homophobia, racism, or other hate speech.\\
        
        \midrule
        Politics & 
        No politics or talking about the election.\\
        
        \midrule
        Formatting & 
        Tag your post, include specific text or information either in the post or comments.\\
        
        \midrule
        Doxxing & 
        No doxxing or posting personal information of an individual.\\
        
        \midrule
        Meta/Enforce & 
        Bans are enforced at moderators discretion.\\
        
        \midrule
        Upvotes & 
        Don't upvote posts that do \_\_\_, or do report posts that do \_\_\_.\\
        
        \midrule
        Off-topic & 
        No posts about specific topic, item, or person.
Post must be relevant.\\
        
        \midrule
        Spam & 
        No spam or reposts, check if its been posted before or read the archives or library for similar posts first.\\
        
        \midrule
        NSFW & 
        No nsfl or nsfw content.
No gore.\\
        
        \midrule
        Other & 
        Microsoft Monday Violation.
See above.
Discord link.\\
        \bottomrule
    \end{tabular}
    \caption{Subreddit rule categories used in the automated labeling process, accompanied by representative examples drawn from the historical dataset.}
    \label{tab:rule_examples}
\end{table*}

\begin{table}[tb]
    \centering
    \small
    \begin{tabular}{lcc}
        \hline
        \textbf{Model / measure} & \textbf{Accuracy} & \textbf{Agreement} \\
        \hline
        Claude Sonnet 4.5 & 67.00\% & Cohen's $\kappa = 0.620$ \\
        GPT-4o            & 66.00\% & Cohen's $\kappa = 0.618$ \\
        Llama 4           & 64.00\% & Cohen's $\kappa = 0.593$ \\
        \hline
        \multicolumn{3}{l}{\textit{Human inter-rater reliability ($N=100$ rules; three coders)}} \\
        Mean pairwise Cohen's $\kappa$ & --- & 0.654 \\
        Fleiss' $\kappa$ & --- & 0.653 \\
        Nominal Krippendorff's $\alpha$ & --- & 0.654 \\
        \hline
    \end{tabular}
    \caption{Performance of automated labeling models compared to human
    consensus, together with human inter-rater reliability measures. Cohen's
    $\kappa$ is reported separately for each model--consensus comparison;
    the human Cohen's $\kappa$ is the mean across the three coder pairs.}
    \label{tab:model_validation}
\end{table}

\subsection{Labeling Prompt}
\begin{quote}
I will give you a phrase from the front page of a subreddit. 
I'm looking for rules for the subreddit. 
If the phrase seems like a typo, then please output 'ERROR'. 
Otherwise, using the following categories, please tell me which, if any of the categories the phrase fits under. 
The categories will be given as a list with at least one example each. 
Output just the number of the category, do not explain or give any other text, do not explain your thinking, do not discuss the result. Output just a number, with no text at all.
Keep in mind the context - all of these phrases come from the side box of the front page of a subreddit.
\end{quote}

The same prompt was given to human coders.

\subsection{Governance Regimes}
\begin{figure*}[t!]
\centering
\includegraphics[width=\textwidth]{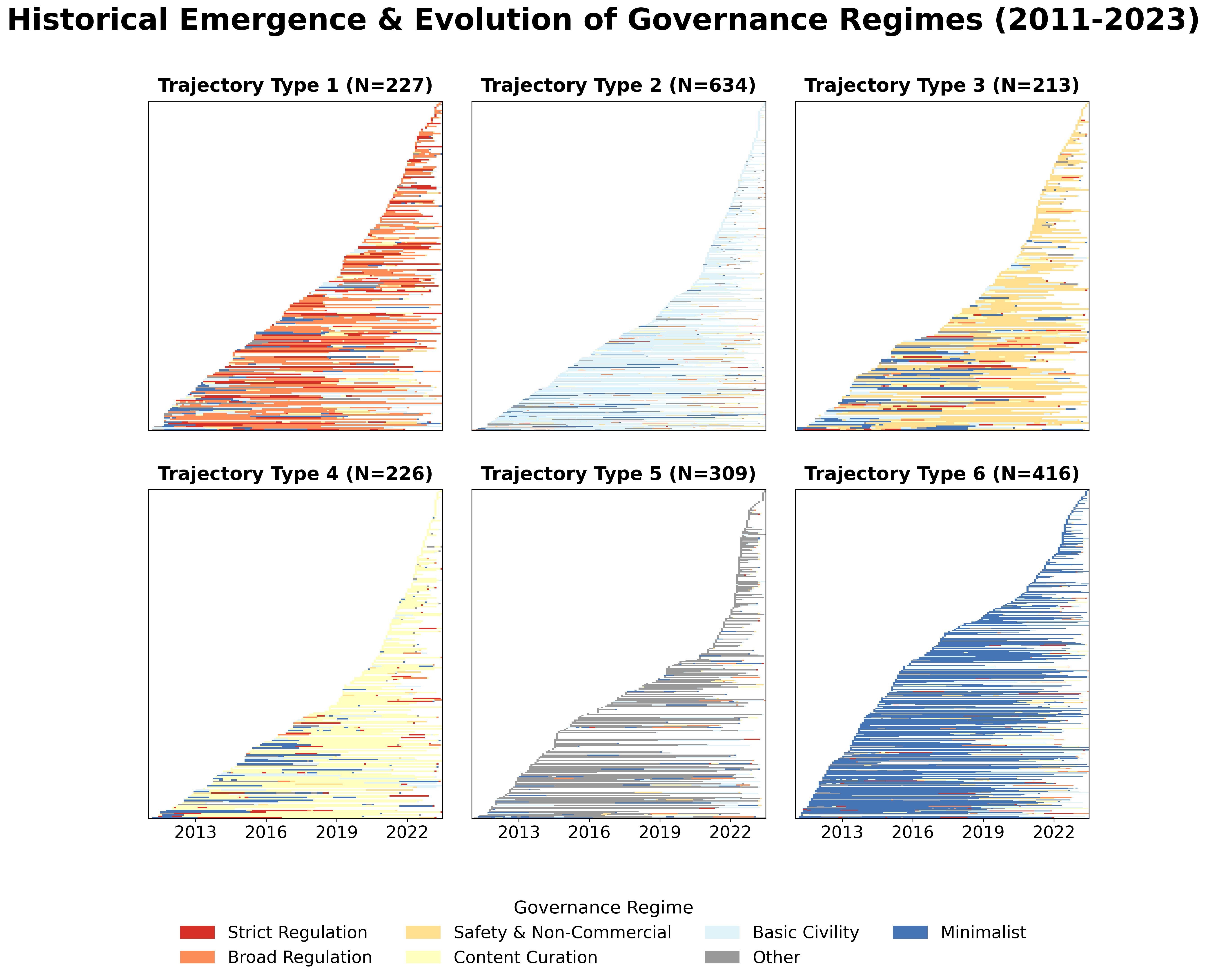}
\caption{Path Dependence of Governance Regimes. Sequence index plots separated by trajectory cluster of regime. Each horizontal line represents one subreddit history, sorted by birth date.}
\label{fig:path_dependence}
\end{figure*}

Figure~\ref{fig:path_dependence} presents the full path-dependence results for all six trajectory clusters, which the main text summarizes but does not display in full. Each horizontal line represents one subreddit's complete regime history from its first archived snapshot through July 2023, sorted within each panel by birth year so that founding-cohort effects are visible as diagonal patterns. The color coding follows the seven-regime scheme from Figure~\ref{fig:regime_def} in the main text.
Several patterns reward close inspection. Trajectory Type 2 (N=634), the largest cluster, is dominated by Basic Civility with occasional excursions into adjacent regimes --- this is the modal Reddit subreddit, one that establishes a harassment norm early and rarely departs from it. Trajectory Types 1 and 3 contain the most visually complex histories, with the most frequent regime switching (about two switches per subreddit, versus roughly one in the other types) --- these are the communities driving the macro-level shock patterns documented in Figure~\ref{fig:appendix_monthly_shocks} below. The sequence plots also make the calcification hypothesis visually intuitive: older subreddits, appearing in the lower rows of each panel given the birth-year sort, show longer unbroken horizontal bands of stable color, while younger ones show more variation. This visual impression is what the dyadic age coefficients in Table~\ref{tab:dyad_attributes} formalize statistically.

\subsection{Shocks \label{sec:shocks}}
\begin{figure*}[t]
\centering
\includegraphics[width=\linewidth]{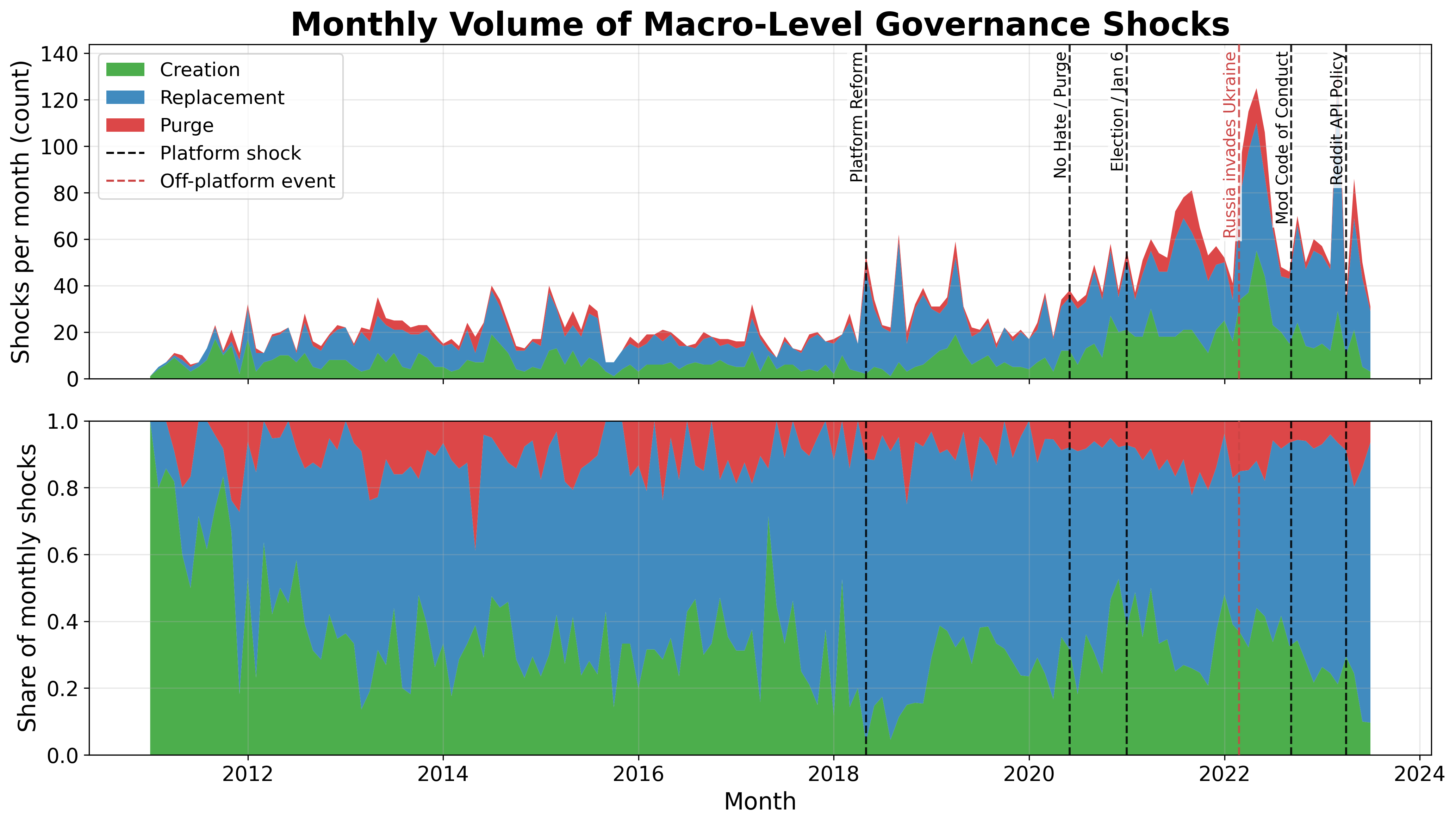}
\caption{Monthly volume (top) and composition share (bottom) of
macro-level governance shocks, 2011--2023. Dashed lines mark
platform-policy events (black) and one off-platform event
(red, the February 2022 Russia--Ukraine invasion). Replacement
events dominate the long run; spikes cluster around the 2018
site redesign, the 2020 hate-speech purge, the early-2022 period
following the Russia--Ukraine invasion and the September 2022
Moderator Code of Conduct, and the early-2023 lead-up to Reddit's
API policy change.}
\label{fig:appendix_monthly_shocks}
\end{figure*}

The shock analysis in the main text reports aggregate counts and highlights three major platform events. The figures in this section provide the full temporal and distributional detail underlying those claims.
Figure~\ref{fig:appendix_monthly_shocks} plots the monthly volume and composition of all 4,717 macro-level governance shocks from 2011--2023. Several features deserve attention beyond what the main text discusses. First, the secular upward trend in total shocks is itself a finding: the rate of constitutional rewrites is increasing over time, consistent with a platform whose governance complexity is compounding. Second, the composition panel confirms that Replacements scale almost linearly with overall volume --- the platform-level spikes are not driven disproportionately by Creations or Purges. Third, the Russia--Ukraine spike in early 2022, marked in red as an off-platform event, is notable precisely because it is not a Reddit policy change: it demonstrates that external societal shocks can trigger governance reorganization independently of platform-internal coercion, a scope condition the main text's H4 framing does not fully capture and which future work should address.

\subsection{Replacements}
\begin{figure}[t]
    \centering
    \includegraphics[width=\linewidth]{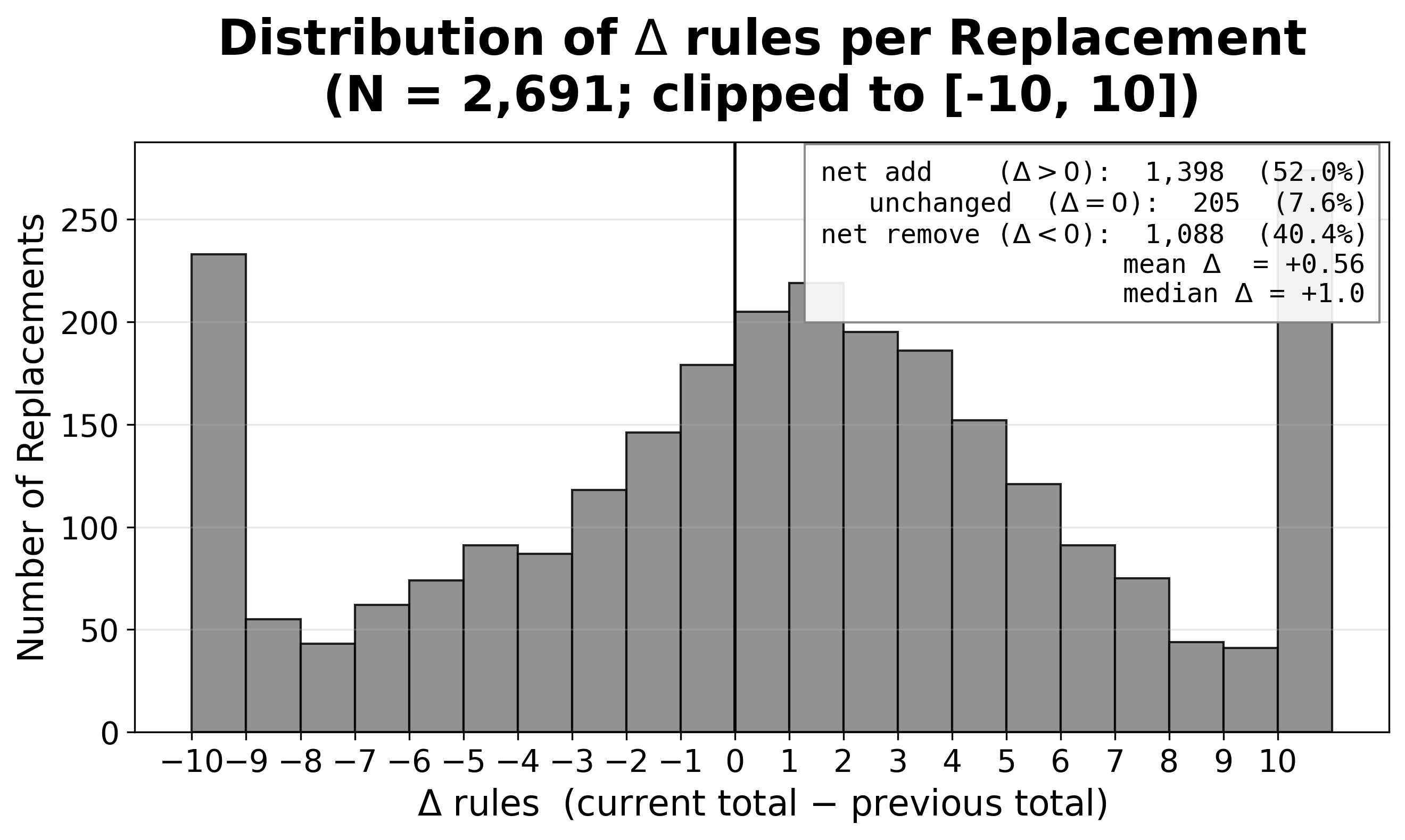}
    \caption{Distribution of $\Delta$ rules (current total $-$ previous total) across all 2{,}691 Replacement events. The mass is sharply concentrated near zero, with $\Delta = +1$ as the modal outcome and a heavy positive tail. Counts and percentages are annotated in the upper-right panel; values are clipped to $[-10, 10]$ for visualisation but full extremes appear in Figure~\ref{fig:appendix_delta_by_year}.}
\label{fig:appendix_delta_hist}
\end{figure}

\begin{figure}[t]
    \centering
    \includegraphics[width=\linewidth]{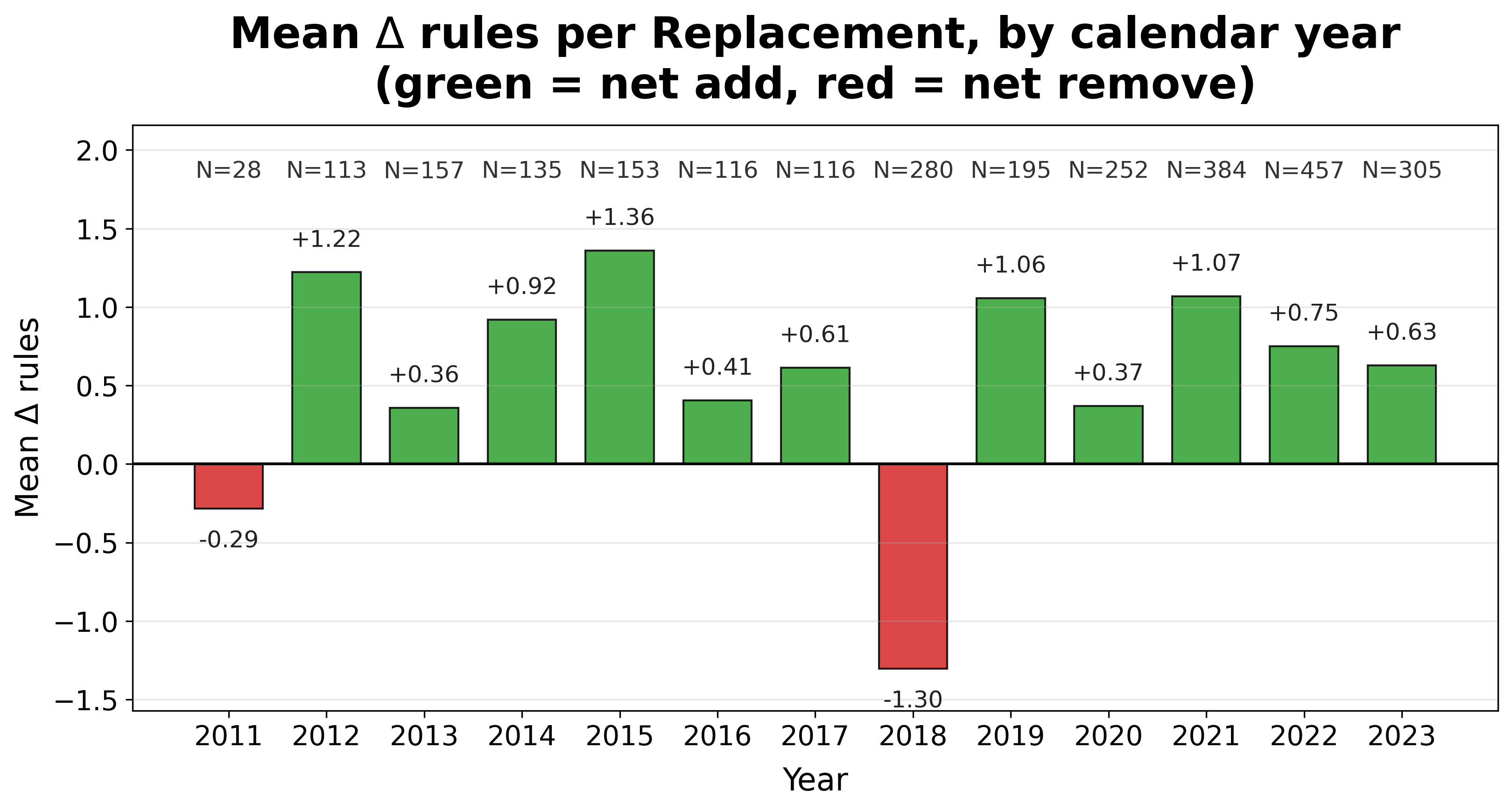}
    \caption{Mean $\Delta$ rules per Replacement event, by calendar year. Green bars denote net-additive years (mean $\Delta > 0$); red bars denote net-removing years (2018, and the low-volume 2011 with $N=28$). The number of Replacement events per year ($N$) is annotated above each bar; the mean $\Delta$ value is annotated at each bar's tip.}
    \label{fig:appendix_delta_by_year}
\end{figure}

Figures~\ref{fig:appendix_delta_hist} and~\ref{fig:appendix_delta_by_year} together provide the distributional and longitudinal evidence for the ratchet interpretation that the main text states but cannot show in full. Figure~\ref{fig:appendix_delta_hist} shows the full distribution of $\Delta$ rules across all 2,691 Replacement events. The modal outcome is $\Delta$=+1: the single most common thing a community does when it rewrites its rulebook is add exactly one rule. The distribution is right-skewed, with a heavy positive tail representing communities that undergo wholesale expansions. The 7.6\% of Replacements with $\Delta$=0 are theoretically interesting in their own right: these communities rewrote their rulebook without changing its size, substituting rules rather than accumulating them. This is the clearest signature of genuine constitutional reform --- ideological rather than structural governance change --- and these events may warrant separate analysis in future work.

Figure~\ref{fig:appendix_delta_by_year} plots mean $\Delta$ rules per Replacement by calendar year. The ratchet interpretation rests heavily on this figure: in every year except 2018 and the low-volume 2011 ($N=28$, mean $\Delta=-0.29$), Replacements added rules on average, with the largest expansions in 2015 (+1.36) and 2021 (+1.07). The 2018 exception (mean $\Delta$=-1.30 across 280 events) is the single most important qualification to the ratchet claim in the paper. It suggests that the 2018 redesign functioned not just as a trigger for governance formalization --- as the shock coefficients in Table~\ref{tab:dyad_shocks} show --- but simultaneously as an occasion for communities to shed legacy rules that no longer fit the new interface or platform norms. The ratchet has a release valve, but opening it required a platform-wide structural intervention, and it re-engaged immediately afterward.

\begin{figure}[t]
    \centering
    \includegraphics[width=\linewidth]{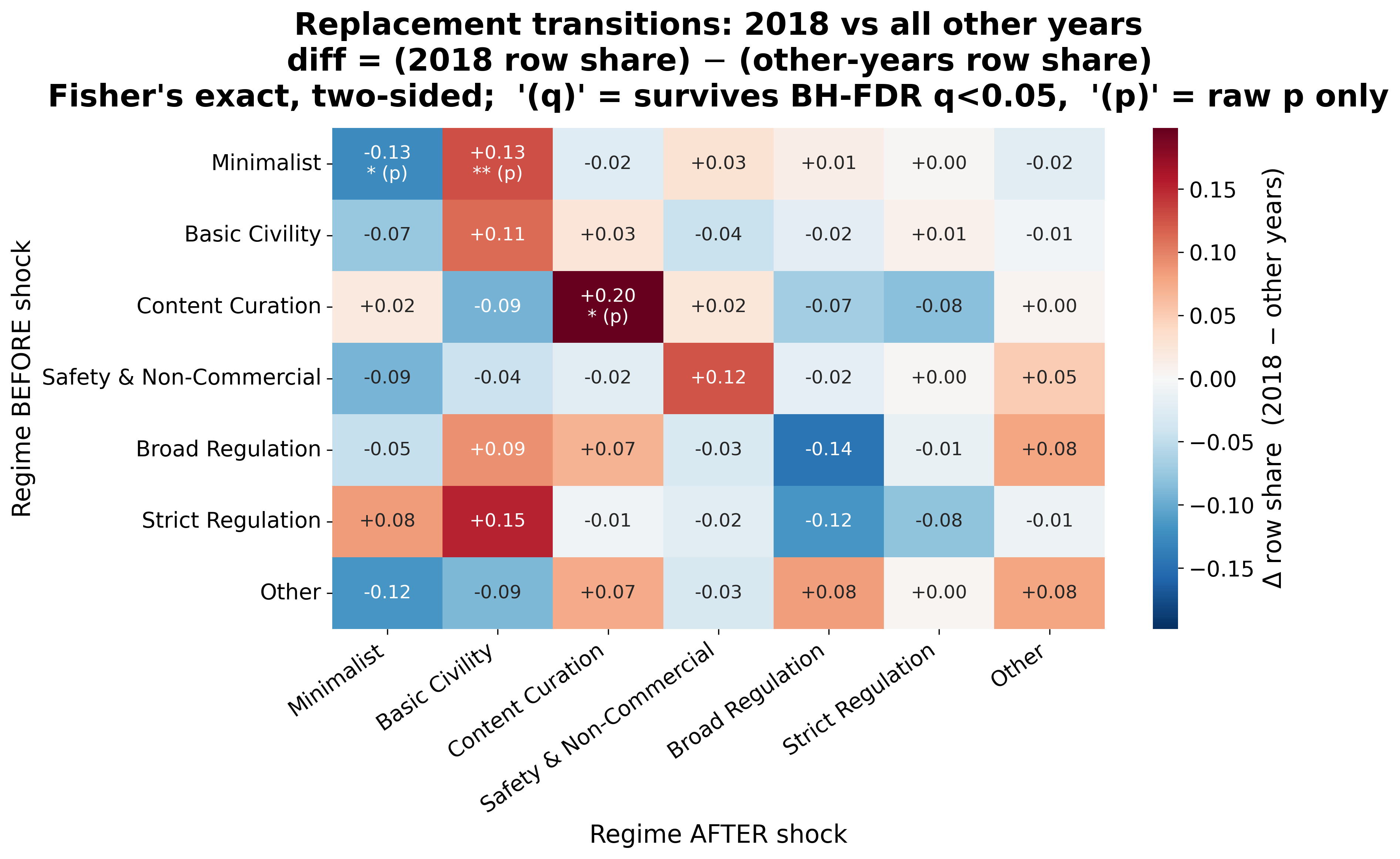}
    \caption{Difference in regime-transition row shares for Replacement events, 2018 versus all other years. Each cell shows (row share in 2018) $-$ (row share in all other years) for the transition $\text{Regime}_{t-1} \to \text{Regime}_t$. Red cells mean the transition was more common in 2018; blue cells mean it was rarer. Cells marked ``(p)'' are significant at raw $p < 0.05$ (Fisher's exact, two-sided); cells marked ``(q)'' survive Benjamini--Hochberg correction at FDR $q < 0.05$ across all 49 tested cells.}
    \label{fig:appendix_2018_diff}
\end{figure}

Figure~\ref{fig:appendix_2018_diff} disaggregates the 2018 shock by showing the cell-by-cell difference in regime-transition row shares between 2018 Replacement events (N=279) and all other years pooled (N=2,333).  Three cells differ at raw $p<0.05$: in 2018, Minimalist communities undergoing a Replacement were less likely to remain Minimalist ($-0.13$, $p=0.030$) and more likely to move into Basic Civility ($+0.13$, $p=0.010$), and Content Curation communities were more likely to remain in Content Curation ($+0.20$, $p=0.034$). None of these survive Benjamini--Hochberg correction across the 49 tested cells, so we treat them as suggestive only. Taken at face value, they point to the redesign prompting formalization at the bottom of the governance ladder. Downward moves from the heaviest regimes (e.g., Strict $\rightarrow$ Basic Civility, $+0.15$) were also somewhat more common in 2018, but not significantly so, and the transition data alone cannot establish where the net rule-shedding of 2018 (Figure~\ref{fig:appendix_delta_by_year}) is concentrated.

\subsection{The Dynamics of Regime Change: Moving Between Broad and Strict Regulation}
\begin{figure*}
    \centering
    \includegraphics[width=\linewidth]{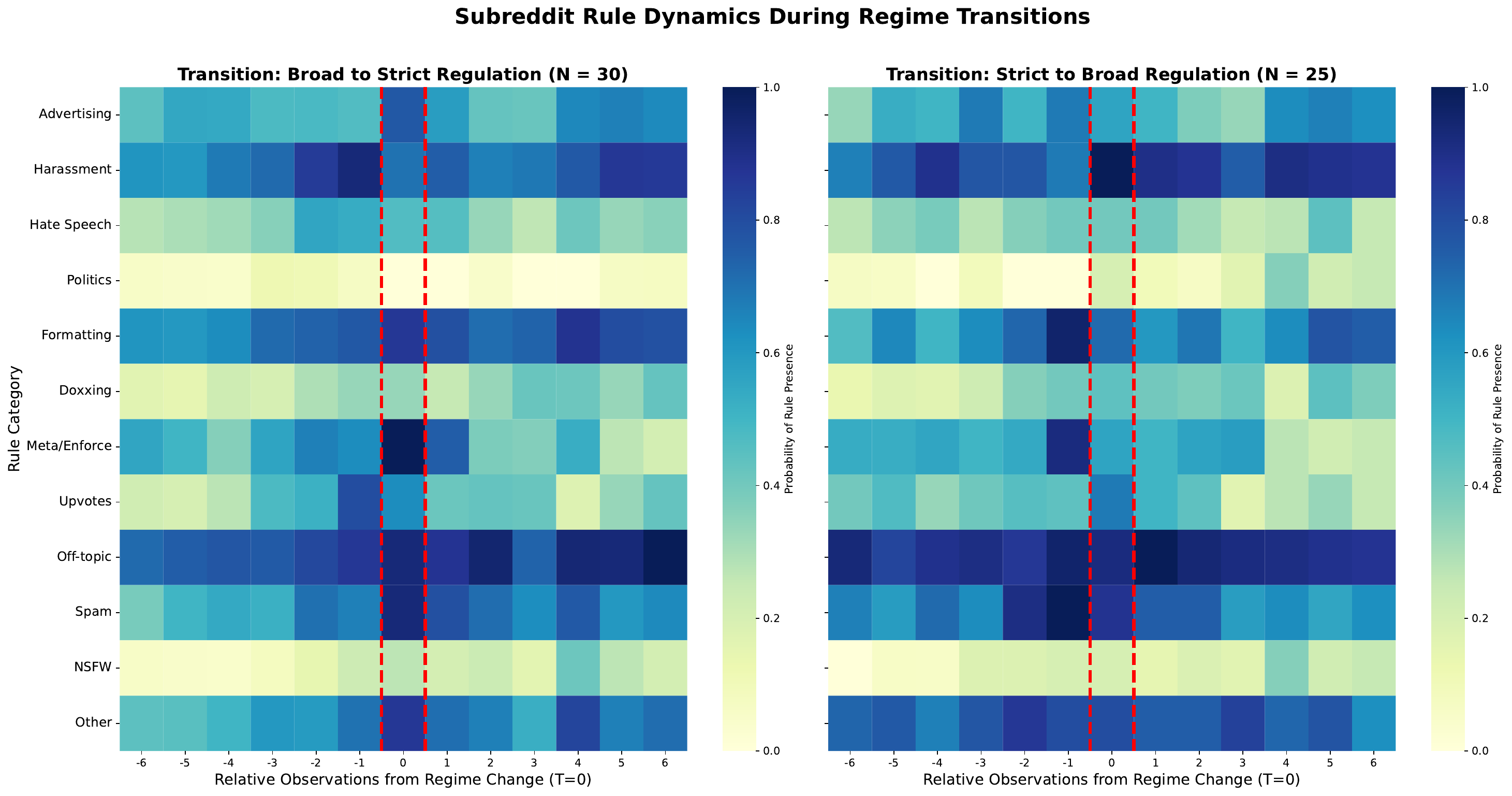}
    \caption{Comparative heatmaps detailing the probability of rule presence before, during, and after a regime change. The left panel shows the escalation from Broad to Strict Regulation ($N=30$), while the right panel shows transitions from Strict back to Broad Regulation ($N=25$). T=0 (bounded by red dashed lines) is the snapshot at which the community was reclassified; the horizontal axis counts archived snapshots relative to that point.}
    \label{fig:regime_transitions}
\end{figure*}

We also examine the temporal dynamics of how communities transition between regimes. Do subreddits slowly drift toward stricter governance, or does bureaucratization occur in sudden bursts? To answer this, we conduct an event study centered on the snapshot at which a community is reclassified between two major states: from \textit{Broad Regulation} to \textit{Strict Regulation} ($N=30$ events), and the reverse ($N=25$).
Figure \ref{fig:regime_transitions} visualizes the probability of a subreddit possessing each of the 12 rule categories across a 13-observation window surrounding the regime change (six archived snapshots before and after the transition at T=0). Because archived snapshots are irregularly spaced, the window is measured in observations rather than calendar months; with so few events, individual cells are noisy and the patterns below should be read as descriptive.
The escalation from \textit{Broad} to \textit{Strict Regulation} (Figure \ref{fig:regime_transitions}, left panel) is preceded by gradual build-up rather than a flat baseline: Spam rules rise from 39\% to 67\% of communities over the six preceding snapshots, and Harassment and Upvotes rules also climb. At T=0 the operational rules that define Strict Regulation peak---Meta/Enforcement reaches 100\% and Spam 93\%. Afterwards the picture is mixed. Spam and Formatting rules stay at or above their pre-transition levels (roughly 60--90\%), consistent with a partial ratchet, but Meta/Enforcement falls back to 21\% by the sixth snapshot, below where it started.
The reverse transition, from \textit{Strict} to \textit{Broad Regulation}, shows the mirror image. At T=0 Meta/Enforcement drops from 92\% to 56\% and Formatting from 96\% to 72\%, while Harassment rules rise to 100\%. These declines do not reverse in the following snapshots: Meta/Enforcement holds near 50--58\% for three snapshots and then falls to roughly 25\%. De-escalations therefore look like genuine reductions in procedural machinery rather than temporary deletions that are quickly restored.
Across both panels, the rules that separate the two regimes are operational ones---Meta/Enforcement, Spam, and Formatting---while Off-topic (72--100\%) and Harassment (60--100\%) rules remain common throughout. Because regime membership is itself defined by rule presence, sharp changes at T=0 are partly mechanical; the more informative features are the build-up before escalation and the absence of a rebound after de-escalation.

\subsection{Dyadic Time Event History Analysis Results}
Tables~\ref{tab:dyad_mechanisms}, \ref{tab:dyad_attributes}, and \ref{tab:dyad_shocks} report the complete log-odds coefficients from the dyadic event history models across all 19 regime and rule outcomes. Regime outcomes use the cluster assignments of the LCA solution shown in Figure~\ref{fig:regime_def}. Very large coefficients that are not significant (e.g., Strict Regulation in 2020--2021, where only 50 adoption events occur) reflect sparse cells and near-separation rather than substantive effects. The main text discusses the theoretically central results organized around H1--H4; this section provides the full tables for replication and for readers interested in outcomes the main text does not discuss.
Table~\ref{tab:dyad_mechanisms} reports the three network channel coefficients. Several results not highlighted in the main text are worth noting. The Politics rule shows a very large negative Shared Moderators coefficient ($\beta = -4.46$) that is not statistically significant; with only 114 adoption events, it most likely reflects sparse data and should not be read as evidence that moderator networks suppress political rules. The Upvotes rule shows a large negative Topic Alignment coefficient ($\beta = -1.13^{*}$), consistent with the interpretation that communities actively differentiate on voting norms as a marker of identity rather than copying topically similar peers. These negative results are as theoretically informative as the positive ones: diffusion channels are not uniformly facilitative but can function as active barriers for socially sensitive rule categories.

\begin{table*}[t]
\centering
\small
\caption{Network Mechanisms of Rule Diffusion (log-odds coefficients).}
\label{tab:dyad_mechanisms}
\begin{tabular}{lccc}
\hline
\textbf{Outcome} & \textbf{Topic Align.} & \textbf{User Overlap} & \textbf{Shared Mods} \\
\hline
\multicolumn{4}{l}{\textit{Regimes}} \\
Broad Regulation        & $0.93$ & $-0.38^{*}$ & $0.01$ \\
Strict Regulation       & $0.88$ & $0.03$ & $1.12$ \\
Content Curation        & $2.45^{***}$ & $0.18$ & $-1.25$ \\
Safety \& Non-Comm.     & $0.32$ & $0.01$ & $-0.51$ \\
Basic Civility          & $1.04^{**}$ & $-0.45^{*}$ & $-0.61$ \\
Minimalist              & $1.07^{*}$ & $-0.30$ & $0.59^{*}$ \\
Other (Regime)          & $-0.51$ & $-1.51^{*}$ & $-0.15$ \\
\hline
\multicolumn{4}{l}{\textit{Rules}} \\
Off-topic               & $1.47^{***}$ & $-0.06$ & $0.82^{***}$ \\
Spam                    & $1.18^{***}$ & $-0.30^{***}$ & $-0.17$ \\
Formatting              & $0.52^{*}$ & $0.05$ & $0.14$ \\
Other (Rule)            & $0.40^{*}$ & $-0.31^{***}$ & $0.83^{***}$ \\
Politics                & $1.11$ & $-0.38$ & $-4.46$ \\
Adv/Comm                & $0.26$ & $0.02$ & $-1.15^{**}$ \\
Mod Pol                 & $-0.44$ & $-0.12$ & $-0.06$ \\
NSFW                    & $-0.18$ & $-0.40$ & $-2.76$ \\
Hate Speech             & $-0.46$ & $-0.32$ & $0.58$ \\
Doxxing                 & $-0.93$ & $-0.48^{*}$ & $-0.20$ \\
Harassment              & $-0.65^{**}$ & $-0.19^{*}$ & $-0.88^{**}$ \\
Upvotes                 & $-1.13^{*}$ & $-0.16$ & $-0.07$ \\
\hline
\multicolumn{4}{p{0.95\columnwidth}}{\footnotesize $^{*}p<.05$, $^{**}p<.01$, $^{***}p<.001$.  Topic Alignment is cosine similarity of \texttt{text-embedding-3-large} description vectors; User Overlap is $\log$ of shared active users; Shared Mods is a binary indicator of overlapping moderator teams.} \\
\end{tabular}
\end{table*}
Table~\ref{tab:dyad_attributes} reports the size and age coefficients. The uniformity of the Target Size effect --- positive and significant for every single outcome ($p<.001$ for all but Strict Regulation, $p<.01$) --- deserves emphasis beyond the main text's H2 discussion. This is not a situation where scale matters for some rule types but not others; it is a universal functional pressure cutting across the entire governance space. The asymmetry between Target and Source size (Source is essentially null across outcomes; the only positive significant coefficient is Doxxing at $\beta = 0.18^{**}$, and the small negative coefficients for Off-topic and Content Curation run opposite to a prestige effect) is equally important and has direct implications for platform design: large subreddits are not governance broadcasters. They adopt rules because they need them, not because smaller communities imitate them. If you want to improve governance quality at scale, you cannot rely on a few large prestigious communities to set standards that propagate downward --- the diffusion network is lateral, not hierarchical.
The age coefficients tell a more nuanced story. Target Age is negative for the most bureaucratic individual rules (Mod Pol, Upvotes, Hate Speech, Spam, Other-Rule), consistent with calcification, and positive for Off-topic. At the regime level the signs are mixed: negative for Broad Regulation and Safety \& Non-Commercial, positive for Strict Regulation, Content Curation, and Basic Civility, and null for Minimalist and Other. Calcification is therefore clearest for the adoption of specific procedural rules rather than for movement between regimes.
\begin{table*}[t]
\centering
\small
\caption{Community Attributes and Rule Adoption (log-odds coefficients).}
\label{tab:dyad_attributes}
\begin{tabular}{lcccc}
\hline
\textbf{Outcome} & \textbf{Tgt.\ Size} & \textbf{Src.\ Size} & \textbf{Tgt.\ Age} & \textbf{Src.\ Age} \\
\hline
Broad Regulation        & $0.50^{***}$ & $0.00$ & $-0.012^{***}$ & $0.000$ \\
Strict Regulation       & $0.27^{**}$ & $-0.06$ & $0.011^{*}$ & $0.003$ \\
Content Curation        & $0.27^{***}$ & $-0.09^{*}$ & $0.004^{*}$ & $0.004^{*}$ \\
Safety \& Non-Comm.     & $0.47^{***}$ & $-0.01$ & $-0.005^{**}$ & $-0.001$ \\
Basic Civility          & $0.20^{***}$ & $0.00$ & $0.004^{***}$ & $0.001$ \\
Minimalist              & $0.27^{***}$ & $0.03$ & $-0.004$ & $-0.001$ \\
Other (Regime)          & $0.32^{***}$ & $-0.05$ & $-0.005$ & $-0.003$ \\
Off-topic               & $0.52^{***}$ & $-0.04^{*}$ & $0.006^{***}$ & $0.000$ \\
Spam                    & $0.46^{***}$ & $0.00$ & $-0.005^{***}$ & $0.001$ \\
Formatting              & $0.55^{***}$ & $-0.01$ & $-0.001$ & $0.000$ \\
Other (Rule)            & $0.35^{***}$ & $0.00$ & $-0.003^{***}$ & $-0.001$ \\
Politics                & $0.51^{***}$ & $-0.02$ & $-0.005$ & $-0.005$ \\
Adv/Comm                & $0.37^{***}$ & $0.01$ & $0.000$ & $0.001$ \\
Mod Pol                 & $0.55^{***}$ & $-0.05$ & $-0.008^{***}$ & $0.001$ \\
NSFW                    & $0.49^{***}$ & $-0.01$ & $-0.002$ & $-0.001$ \\
Hate Speech             & $0.49^{***}$ & $-0.07$ & $-0.007^{***}$ & $0.001$ \\
Doxxing                 & $0.36^{***}$ & $0.18^{**}$ & $0.001$ & $0.001$ \\
Harassment              & $0.25^{***}$ & $0.02$ & $0.000$ & $0.002^{*}$ \\
Upvotes                 & $0.47^{***}$ & $0.05$ & $-0.013^{***}$ & $0.001$ \\
\hline
\multicolumn{5}{p{0.95\columnwidth}}{\footnotesize $^{*}p<.05$, $^{**}p<.01$, $^{***}p<.001$.  Sizes in $\log$ active users; ages in months.} \\
\end{tabular}
\end{table*}
Table~\ref{tab:dyad_shocks} reports the platform-intervention period effects. The full table reveals several additional results. The 2018 Reform simultaneously increased Politics ($\beta = 2.10^{***}$) and Upvotes ($\beta = 2.32^{***}$) while suppressing Off-topic ($\beta = -0.73^{***}$, Formatting ($\beta = -1.07^{***}$), and Mod Pol ($\beta = -0.87^{***}$) --- suggesting the redesign redirected governance attention toward participation norms and away from content curation and procedural machinery, possibly because the new interface rendered some of that machinery redundant. The 2020 Deplatforming period's suppression of Hate Speech peer-copying $\beta = -1.30^{***}$) and Harassment ($\beta = -1.04^{***}$) is the clearest illustration of the displacement mechanism discussed in the main text: when the platform absorbs a regulatory burden directly, local communities stop importing that rule from each other. The 2021 Election period's strong negative effects on NSFW ($\beta = -1.06^{***}$), Advertising ($\beta = -0.36^{**}$), and the Safety \& Non-Commercial regime ($\beta = -1.47^{***}$), with no regime showing a significant increase, point to a general slowdown in peer-to-peer adoption during this period rather than consolidation around particular regimes.
\begin{table*}[t]
\centering
\small
\caption{Policy-Era Effects on Dyadic Adoption (log-odds coefficients).}
\label{tab:dyad_shocks}
\begin{tabular}{lccc}
\hline
\textbf{Outcome} & \textbf{Reform 2018} & \textbf{Deplatforming 2020} & \textbf{Election 2021} \\
\hline
Broad Regulation        & $1.69^{***}$ & $-0.14$ & $-0.16$ \\
Strict Regulation       & $0.63$ & $-8.13$ & $8.33$ \\
Content Curation        & $-0.63$ & $-0.53$ & $0.50$ \\
Safety \& Non-Comm.     & $0.13$ & $1.29^{***}$ & $-1.47^{***}$ \\
Basic Civility          & $1.29^{***}$ & $0.91^{***}$ & $0.22$ \\
Minimalist              & $-1.30^{***}$ & $-4.08^{*}$ & $3.20$ \\
Other (Regime)          & $0.34$ & $0.24$ & $-1.36^{*}$ \\
Off-topic               & $-0.73^{***}$ & $-0.77^{***}$ & $0.06$ \\
Spam                    & $0.48^{*}$ & $0.45^{***}$ & $-0.01$ \\
Formatting              & $-1.07^{***}$ & $-0.58^{***}$ & $-0.49^{***}$ \\
Other (Rule)            & $-0.12$ & $-0.28^{**}$ & $-0.45^{***}$ \\
Politics                & $2.10^{***}$ & $0.34$ & $-0.15$ \\
Adv/Comm                & $1.14^{***}$ & $0.13$ & $-0.36^{**}$ \\
Mod Pol                 & $-0.87^{***}$ & $-0.10$ & $0.22$ \\
NSFW                    & $-0.11$ & $-0.19$ & $-1.06^{***}$ \\
Hate Speech             & $0.18$ & $-1.30^{***}$ & $0.33$ \\
Doxxing                 & $0.86^{*}$ & $-6.37$ & $4.47$ \\
Harassment              & $0.33^{*}$ & $-1.04^{***}$ & $0.03$ \\
Upvotes                 & $2.32^{***}$ & $-4.64$ & $3.88$ \\
\hline
\multicolumn{4}{p{0.95\columnwidth}}{\footnotesize $^{*}p<.05$, $^{**}p<.01$, $^{***}p<.001$.  Reform 2018: May 2018 site overhaul.  Deplatforming 2020: June 2020 hate-speech policy + r/The\_Donald ban.  Election 2021: Jan 2021 post-election period.} \\
\end{tabular}
\end{table*}

\subsection{Niches}
\begin{figure*}[htbp]
    \centering
    \includegraphics[width=\textwidth]{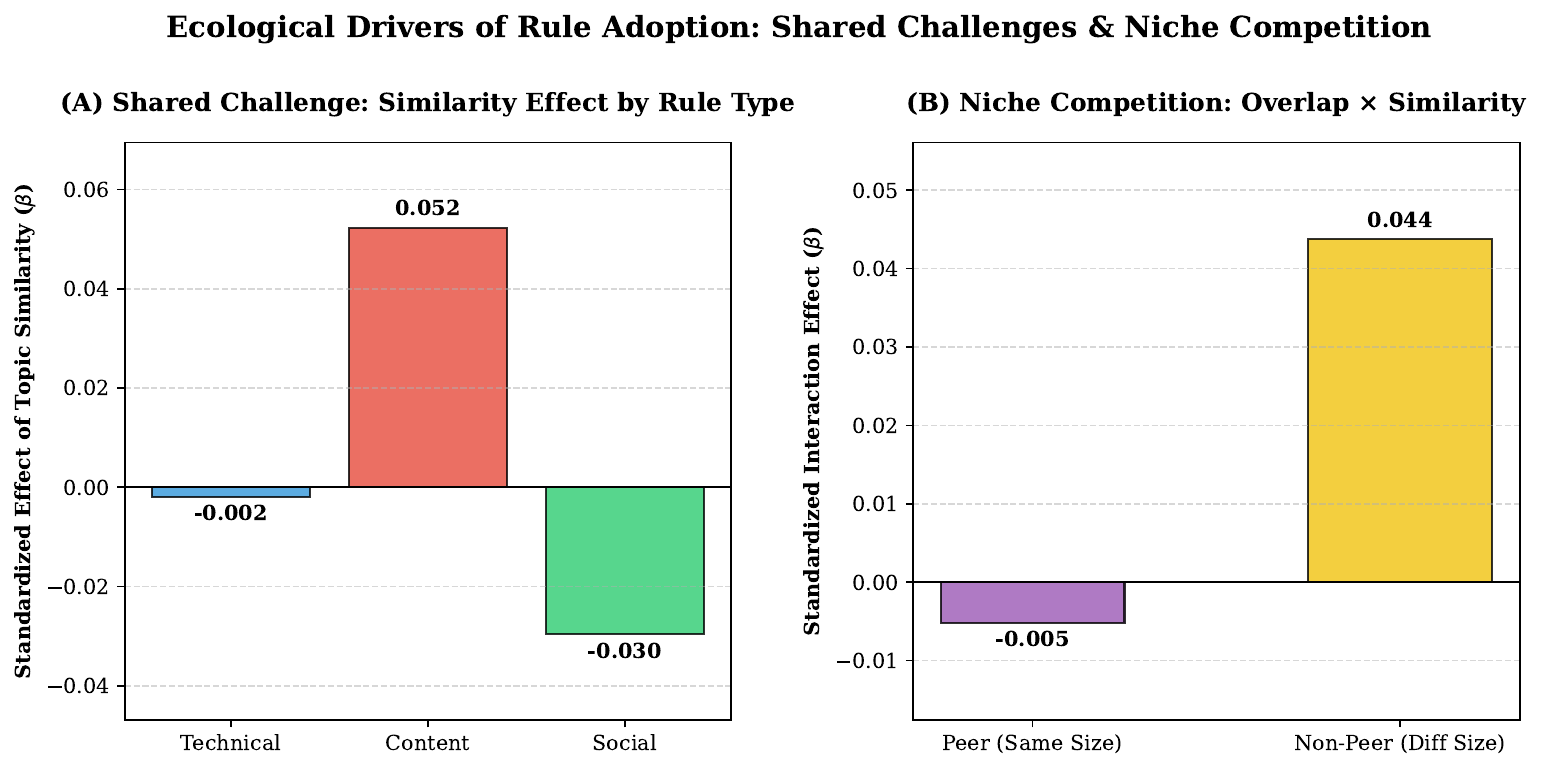}
    \caption{\textbf{Ecological Drivers of Rule Adoption: Shared
Challenges \& Niche Competition}. (A) Standardized main effect
of topic similarity ($\beta$) across rule categories. Controlling
for structural factors reveals that similarity strictly drives the adoption of Content rules ($\beta = 0.052$), while Technical and Social rules face adoption resistance ($\beta = -0.002$ and $\beta = -0.030$, respectively). (B) Standardized interaction effect between user overlap and topic similarity. Similarity-driven diffusion is robust among non-peers occupying different niches ($\beta = 0.044$) but is fully inhibited by differentiation pressure
among structural peers occupying the same size niche ($\beta =
-0.005$).}
    \label{fig:ecological_diffusion}
\end{figure*}

Figure~\ref{fig:ecological_diffusion} presents the two-panel ecological drivers analysis that tests the shared challenge and niche competition hypotheses developed in the main text. Both panels plot standardized effect sizes, allowing direct magnitude comparison across rule categories and structural groups.
Panel A plots the standardized main effect of topic similarity ($\beta$) across the three aggregated rule categories (Technical, Content, Social), controlling for structural confounds including source size, target size, community age, global rule prevalence, and time-period fixed effects. The stark divergence --- Content rules at +0.052, Technical at -0.002, Social at -0.030 --- is the empirical core of the shared challenge hypothesis: communities copy each other's content governance when they face similar moderation challenges defined by their topical niche, but actively resist copying operational and behavioral rules from topically similar peers. The resistance for Social rules suggests that communities treat behavioral norms --- how to handle harassment, what commercial activity is acceptable --- as markers of community culture and identity, differentiating on them even from communities that face objectively similar social challenges. Behavioral governance is not a solution to a shared problem; it is a statement of who a community is.
Panel B plots the interaction of user overlap and topic similarity split by structural peer status (same-size versus different-size communities), introduced to test whether niche competition inhibits similarity-driven diffusion. The sign reversal --- positive interaction among non-peers (+0.044), neutral-to-negative among peers (-0.005) --- isolates the differentiation inhibition mechanism. Communities that share a topic and share users but occupy different size niches treat each other as complementary and copy freely; the same communities at the same size treat each other as competitors and resist convergence. This finding connects the diffusion analysis directly back to the ecological framework in the related work: Carroll's resource-partitioning logic operates not just at the population level but at the dyadic level within a single platform, and it operates on governance --- not just on content or audience --- as a domain of competitive differentiation.

\end{document}